\documentclass[letterpaper, 10 pt, conference]{ieeeconf}  

\IEEEoverridecommandlockouts                              

\usepackage{graphics} 
\usepackage{graphicx}
\usepackage{amsmath} 
\usepackage{amssymb}  
\usepackage{algorithmic, algorithm}
\usepackage{float}
\usepackage{multirow}

\newtheorem{definition}{Definition}

\newcommand{\figsize}{0.225}

\title{\LARGE \bf
A Novel Multivariate Skew-Normal Mixture Model and Its Application in
Path-Planning for Very-Large-Scale Robotic Systems 
}

\author{Pingping Zhu$^{1}$, Chang Liu$^{2}$ and Peter Estephan$^{3}$
	\thanks{*This research was made possible
		by the NASA Established Program to Stimulate Competitive Research, Grant \# 80NSSC22M0027 and John Marshall University Scholars Awards}
	\thanks{$^{1}$Pingping Zhu is with Faculty of Computer Sciences and Electrical Engineering, Marshall University, Huntington, WV, USA
		        {\tt\small zhup@marshall.edu}}%
	\thanks{$^{2}$Chang Liu is with the Department of Advanced Manufacturing and Robotics, College of Engineering, Peking University, Beijing 100871, China {\tt\small changliucoe@pku.edu.cn}}		        
	\thanks{$^{3}$Peter Estephan with the Department of Computer Sciences and Electrical Engineering, Marshall University, Huntington, WV, USA
		        {\tt\small estephan@marshall.edu}}%
	}

\begin{document}

\maketitle
\thispagestyle{empty}
\pagestyle{empty}

\begin{abstract}

This paper addresses the path-planning challenge for very large-scale robotic systems (VLSR) operating in complex and cluttered environments. VLSR systems consist of numerous cooperative agents or robots working together autonomously. Traditionally, many approaches for VLSR systems are developed based on Gaussian mixture models (GMMs), where the GMMs represent agents' evolving spatial distribution, serving as a macroscopic view of the system's state. However, our recent research into VLSR systems has unveiled limitations in using GMMs to represent agent distributions, especially in cluttered environments. To overcome these limitations, we propose a novel model called the skew-normal mixture model (SNMM) for representing agent distributions. Additionally, we present a parameter learning algorithm designed to estimate the SNMM's parameters using sample data. Furthermore, we develop two SNMM-based path-planning algorithms to guide VLSR systems through complex and cluttered environments. Our simulation results demonstrate the effectiveness and superiority of these algorithms compared to GMM-based path-planning methods.

\end{abstract}

\section{INTRODUCTION}
With rapid advancements in computing and sensor technology, there is a growing interest in very-large-scale robotic (VLSR) systems. These systems, composed of numerous cooperative robots, offer significant advantages over single or small groups of robots, especially when they collaborate on common tasks through information sharing \cite{steinberg2016long, huttenrauch2019deep, hernandez2019survey}.

Various intelligent control approaches have emerged to understand and control the collective behavior of VLSR systems, particularly in scenarios like guiding or tracking robotic swarms through obstacle-filled environments. Because the optimal planning for cooperative robots is PSPACE-hard \cite{parker2009path}, the scalability issue is challenging  for the VLSR problems. One promising approach is distributed optimal control (DOC) \cite{foderaro2014distributed, ferrari2016distributed, rudd2017generalized, zhu2021adaptive}, which uses time-varying probability density functions (PDFs) of robots or agents to represent the system's macroscopic states. These PDFs guide the microscopic control of robots, addressing scalability challenges in VLSR systems. The Gaussian mixture model (GMM) is often used to model macroscopic states due to its simplicity and representation capabilities. However, recent research has revealed GMM's limitations, especially in cluttered environments with many small-scale obstacles, such as forests \cite{soria2021distributed}. 
This limitation of the GMM arises from its components, Gaussian or normal distributions, which are widely used in scientific and engineering areas due to their simple mathematical expression. However, Gaussian distributions' simplicity reduces their flexibility and adaptability in representing robot distributions against complex environments, especially the cluttered environments.  A large number of Gaussian components are required to construct a GMM to obtain an acceptable distribution representation performance, which violates the GMMs' simplicity advantage.

This paper proposes novel statistical models based on multivariate skew-normal (SN) distributions to represent agent distributions in complex environments better. The univariate skew-normal family was formally defined by Azzalini in 1985 \cite{azzalini1985skewNormal}, and subsequent research introduced various definitions of multivariate skew-normal distributions \cite{azzalini1996multivariate, gupta2004multivariate, arellano2005fundamental}. These distributions share a crucial feature: their PDFs are proportional to the product of a Gaussian PDF and a skewing function, where the skewing function is typically a probability function or cumulative density function (CDF). Our approach considers interactions between agent distributions and obstacles, treating the obstacles' occupancy probability function \cite{ramos2016hilbert} as the skewing function. Additionally, we introduce the skew-normal mixture model (SNMM) to represent VLSR system macroscopic states, extending our skew-normal distributions. We also present a parameter learning algorithm to estimate SNMM parameters from agent positions. Simulation results demonstrate that this algorithm accurately estimates SNMM parameters, offering a superior macroscopic representation of agents with fewer components in cluttered, obstacle-filled environments than GMM. 
To showcase SNMM's applications in VLSR systems, finally, we develop two SNMM-based path-planning algorithms for guiding large groups of agents through complex and cluttered environments,  e.g., artificial forests. Simulations illustrate the effectiveness and superiority of these SNMM-based algorithms over GMM-based alternatives.




\section{MULTIVARIATE SKEW-NORMAL MIXTURE MODEL}
\label{sec: SNMM}
A fundamental skew-normal (FUSN) distribution is defined in \cite{arellano2005fundamental}, and a modified version is provided as follows:
\begin{definition}[Fundamental Skew-Normal Distribution] \label{def:FSND}
	Consider an $n_z \times 1$ random vector $\mathbf{Z} \in \mathbb{R}^{n_z}$, and 
	an $n_x \times 1$ random vector,  $\mathbf{X} \in \mathbb{R}^{n_x}$, where the random vector $\mathbf{X}$ is multivariate normal  distributed, such that  $\mathbf{X} \sim \mathcal{N}(\boldsymbol{\mu},\boldsymbol{\Sigma})$. The random variable,  $\mathbf{X}^* = \left[ \mathbf{X} \vert \mathbf{Z} < \boldsymbol{\zeta} \right]$, is said to have a fundamental skew-normal (FUSN) distribution, denoted by $\mathbf{X}^* \sim FUSN(\boldsymbol{\mu}, \boldsymbol{\Sigma}, Q_{\mathbf{Z}})$, if its PDF is given by
	\begin{equation}
	f_{\mathbf{X}^*}(\mathbf{x}) = \frac{\phi_{\mathbf{X}} (\mathbf{x} \vert \boldsymbol{\mu}, \boldsymbol{\Sigma}) Q_{\mathbf{Z}}(\mathbf{x})}{\mathbb{E}_{\mathbf{X}\sim \phi_{\mathbf{X}}}[Q_{\mathbf{Z}}(\mathbf{X})]}
	\label{eq: PDF_FUSN}
	\end{equation} 
	where $\boldsymbol{\zeta} \in \mathbb{R}^{n_z}$ is a user-defined threshold vector, $\phi_{\mathbf{X}} (\mathbf{x} \vert \boldsymbol{\mu}, \boldsymbol{\Sigma})$ is the PDF of the multivariate normal random vector $\mathbf{X}$, and $Q_{\mathbf{Z}}(\mathbf{x}) \triangleq P(\mathbf{Z} < \boldsymbol{\zeta} \vert \mathbf{X} = \mathbf{x})$. 
\end{definition}
Here, the function $Q_{\mathbf{Z}}(\mathbf{x})$ is a skewing function,  and $\mathbb{E}_{\mathbf{X}\sim \phi_{\mathbf{X}}}[\cdot]$ denotes the expectation with respect to (w.r.t.) the PDF $\phi_{\mathbf{X}}$. According to \cite{arellano2005fundamental}, \textbf{Definition} \ref{def:FSND} can contain many different families of skew-normal distributions. In this section, we propose a new skew-normal distribution based on  \textbf{Definition} \ref{def:FSND} by embedding the environmental information. Furthermore, a skew-normal mixture model (SNMM) is proposed to describe the statistics of VLSR systems in obstacle-deployed environments. Finally, a parameter learning algorithm is developed to estimate the model parameters of the SNMM from samples.  

\subsection{Multivariate Skew-Normal Distribution}
In the VLSR problems, the position of the $n$th agent can be considered as a continuous random vector, $\mathbf{X}_n \in \mathcal{W}$, $n=1,\ldots, N$, where $\mathcal{W} \subset \mathbb{R}^2$ denotes the workspace. If we assume that $\{\mathbf{X}_n\}_{n=1}^N$ are all independent and identically distributed (\textit{i.i.d.}) multivariate normal random vector denoted by $\mathbf{X} \sim \mathcal{N}(\boldsymbol{\mu}, \boldsymbol{\Sigma})$, then the actual positions of $N$ agents can be treated as $N$ independent samples/realizations generated from $\mathbf{X}$. 

On the other hand,  let $\mathcal{B} \subset \mathcal{W}$ denote the set of the locations occupied by obstacles in the workspace. We can have a Bernoulli random variable $Y(\mathbf{x}) \in \{0, 1\}$ indicating if the location $\mathbf{x} \in \mathcal{W}$ is occupied by the obstacles, such that $Y(\mathbf{x}) = 1$ if $\mathbf{x} \in \mathcal{B}$, otherwise, $Y(\mathbf{x}) = 0$. Thus, $Y(\mathbf{X})$ defined on the whole workspace specifies a Bernoulli-random-field (BRF). Furthermore, consider the scenario where the obstacle detection can be determined via observing a continuous random measurement, $\mathbf{Z}(\mathbf{x})$, obtained at $\mathbf{x}$, such that $P(Y(\mathbf{x}) = 1 ) = P(\mathbf{Z}(\mathbf{x})  \geq \boldsymbol{\zeta} )$, where $\boldsymbol{\zeta}$ indicates a user-defined threshold parameter. Then, because the agents cannot be deployed on the obstacles, the distribution of the agents in an obstacle-deployed environment can be modeled by a FUSN distribution defined in \textbf{Definition} \ref{def:FSND}. 

Finally, considering the direct interaction between agents and obstacles, we can construct a novel skew-normal distribution based on FUSN to describe the statistical model for the VLSR problem referred to as the Bernoulli-random-filed based skew-normal (BRF-SN) distribution, which is defined as follows.
\begin{definition}[Bernoulli-Random-Field based Skew-Normal Distribution] \label{def:BRF-SND}
	Consider a continuous random vector,  $\mathbf{X} \in \mathbb{R}^{n_x}$ and a Bernoulli random variable $Y \in \{0, 1\}$, where the random vector $\mathbf{X}$ is multivariate normal  distributed, such that  $\mathbf{X} \sim \mathcal{N}(\boldsymbol{\mu},\boldsymbol{\Sigma})$. The random variable,  $\mathbf{X}^* = \left[ \mathbf{X} \vert Y = 0 \right]$, has a Bernoulli-random-field based skew-normal (BRF-SN) distribution denoted by $\mathbf{X}^* \sim BRFSN(\boldsymbol{\mu}, \boldsymbol{\Sigma}, Q_Y)$, if its PDF is given by
	\begin{equation}
		f_{\mathbf{X}^*}(\mathbf{x}) = \frac{\phi_{\mathbf{X}} (\mathbf{x} \vert \boldsymbol{\mu}, \boldsymbol{\Sigma}) Q_Y(\mathbf{x})}{\mathbb{E}_{\mathbf{X}\sim \phi_{\mathbf{X}}}[Q_Y(\mathbf{X})]}
		\label{eq: PDF_BRFSN}
	\end{equation} 
	where $\phi_{\mathbf{X}} (\mathbf{x} \vert \boldsymbol{\mu}, \boldsymbol{\Sigma})$ is the PDF of the multivariate normal random vector $\mathbf{X}$, and $Q_Y(\mathbf{x}) \triangleq P(Y(\mathbf{X}) = 0 \vert \mathbf{X} = \mathbf{x})$. 
\end{definition}

Notably, the skewing function, $Q_Y(\mathbf{x})$, can be obtained through many approaches, e.g., the Hilbert occupancy map. The readers are referred  to  \cite{ramos2016hilbert, morelli2019integrated, zhu2019scalable, zhu2021adaptive}. 

\subsection{Skew-Normal Mixture Model}
To describe a more complex agents' distribution, we extended the BRF-SN distribution to a mixture model associated with the following PDF,
\begin{equation}
	\wp_{\mathbf{X}^*} (\mathbf{x}) \triangleq  \sum_{i=1}^{N_C} \omega_i f_{\mathbf{X}^*, i} (\mathbf{x} \vert \boldsymbol{\mu}_i, \boldsymbol{\Sigma}_i) 
	\label{eq: def_BRF_SNMM}
\end{equation}  
This statistical model, referred to as the Bernoulli-random-field based skew-normal mixture model (BRF-SNMM), consists of $N_C$ BRF-SN components, $\{f_{\mathbf{X}^*, i}(\mathbf{x} \vert \boldsymbol{\mu}_i, \boldsymbol{\Sigma}_i)\}_{i=1}^{N_C}$, defined in \textbf{Definition} \ref{def:BRF-SND} associated with the corresponding weights, $\{\omega_i\}_{i=1}^{N_C}$. Given the skewing function $Q_Y(\mathbf{x})$, therefore, the BRF-SNMM is specified by the component parameters $\boldsymbol{\Theta}_i = (\boldsymbol{\mu}_i, \boldsymbol{\Sigma}_i)$, $i = 1,\ldots, N_C$, and the weight parameters, $\boldsymbol{\omega} = \begin{bmatrix}
	\omega_1 & \ldots & \omega_{N_C}
\end{bmatrix}$, such that $\sum_{i=1}^{N_C} \omega_i = 1$ and $\omega_i > 0$ for all $i$, which are also denoted by $\boldsymbol{\Theta}_{1:N_C} = \{ (\omega_i, \boldsymbol{\Theta}_i )\}_{i=1}^{N_C}$ for short.

\subsection{Parameters Learning For Skew-Normal Mixture Model}
\label{subsec: paramLearning_SNMM}
Because the BRF-SNMM has a structure similar to the GMM, we can also use the expectation-maximization (EM) method to learn the parameters from the given samples,  $\mathcal{D} = \{\mathbf{x}_n\}_{n=1}^N$. First, we consider an $N_C$-dimensional categorical random variable, $C \in \{1,\ldots, N_C\}$, indicating the label of the BRF-SN components. Given the parameters, $\boldsymbol{\Theta}_{1:N_C}$, the conditional distribution can be specified by the component weight $\boldsymbol{\omega}$, such that $P(C = i \vert \boldsymbol{\Theta}_{1:N_C}) = \omega_i, \text{ for } i = 1,\ldots, N_C$. 
Thus, the BRF-SN distribution,  $f_{\mathbf{X}^*_i} (\mathbf{x} \vert \boldsymbol{\mu}_i, \boldsymbol{\Sigma}_i)$, can be interpreted as the conditional distribution of the random variable $\mathbf{X}^*$, given $\mathbf{X}^*$ is associated with the $i$th BRF-SN component, such that $\wp_{\mathbf{X}^* \vert C} (\mathbf{x} \vert C = i, \boldsymbol{\Theta}_{1:N_C}) = f_{\mathbf{X}^*, i} (\mathbf{x} \vert \boldsymbol{\mu}_i, \boldsymbol{\Sigma}_i)$.

To specify the unknown underlying parameters, $\boldsymbol{\Theta}_{1:N_C}$, from
the samples, we can construct the following negative loglikelihood
(NLL) object function,  
\begin{equation}
	J(\boldsymbol{\Theta}_{1:N_C} \vert \mathcal{D}) \triangleq \sum_{n=1}^N  \mathcal{L}(\mathbf{x}_n, \boldsymbol{\Theta}_{1:N_C})
	\label{eq:NLL}
\end{equation}
where the term, $\mathcal{L}(\mathbf{x}_n, \boldsymbol{\Theta}_{1:N_C})$, is defined by
\begin{align}
	\mathcal{L}(\mathbf{x}_n, \boldsymbol{\Theta}_{1:N_C}) &\triangleq -\ln \bigg[ \wp_{\mathbf{X}^*} (\mathbf{x}_n \vert \boldsymbol{\Theta}_{1:N_C})  \bigg] \\ 
	&= 	-\ln \mathbb{E}_{C\vert \mathbf{x}_n} \bigg[ \frac{\wp_{\mathbf{X}^*,  C} (\mathbf{x}_n, C = i \vert \boldsymbol{\Theta}_{1:N_C} )}{P(C = i\vert \mathbf{x}_n)} \bigg] \nonumber
\end{align}
Here, $\mathbb{E}_{C\vert \mathbf{x}_n} [ \cdot ]$ indicates the expectation term w.r.t. the discrete distribution $P(C\vert \mathbf{x}_n)$.

Similar to the GMM parameter learning approach, a upper bound of the NLL object function is provided as follows,
\begin{equation}
	\tilde{J}( \boldsymbol{\Gamma}, \boldsymbol{\Theta}_{1:N_C} \vert \mathcal{D}) \triangleq \sum_{n=1}^N \tilde{\mathcal{L}}(\mathbf{x}_n, \boldsymbol{\gamma}_n, \boldsymbol{\Theta}_{1:N_C}) 
	\label{eq: costFunc_paramLearning}
\end{equation} 
where $\gamma_{n,i} \triangleq \wp_{C \vert \mathbf{X}^*} (C 
= i \vert \mathbf{x}_n,  \boldsymbol{\Theta}_{1:N_C} )$, $\boldsymbol{\gamma}_n = \begin{bmatrix}
	\gamma_{n,1} & \ldots & \gamma_{n,N_C}
\end{bmatrix} = \wp_{C \vert \mathbf{X}^*} (C 
\vert \mathbf{x}_n,  \boldsymbol{\Theta}_{1:N_C} )$, and  $\boldsymbol{\Gamma} = \begin{bmatrix}
	\gamma_{n,i}
\end{bmatrix}_{n=1, i=1}^{N, N_C}$ is a $N \times N_C$ matrix. Here, the term   $\tilde{\mathcal{L}}(\mathbf{x}_n, \boldsymbol{\gamma}_n, \boldsymbol{\Theta}_{1:N_C})$  is a tight upper bound of $\mathcal{L}(\mathbf{x}_n, \boldsymbol{\Theta}_{1:N_C})$, which is defined by 
\begin{align}
	\tilde{\mathcal{L}}(\mathbf{x}_n, \boldsymbol{\gamma}_n, \boldsymbol{\Theta}_{1:N_C}) &\triangleq  \sum_{i=1}^{N_C} \gamma_{n,i} \ln \bigg[ \frac{\gamma_{n,i}}{\wp_{\mathbf{X}^*, C} (\mathbf{x}_n, C = i \vert  \boldsymbol{\Theta}_{1:N_C} )} \bigg]\nonumber\\
	&=	\sum_{i=1}^{N_C} \gamma_{n,i} \ln \bigg[ \frac{\gamma_{n,i}}{\omega_i f_{\mathbf{X}^*, i} (\mathbf{x} \vert \boldsymbol{\mu}_i, \boldsymbol{\Sigma}_i) } \bigg]
\end{align}
The upper bound, $\tilde{\mathcal{L}}(\mathbf{x}_n, \boldsymbol{\gamma}_n, \boldsymbol{\Theta}_{1:N_C})$, is obtained according to Jensen's inequality \cite{durrett2019probability}. The details are omitted because of limited pages. 

Then, we apply the EM method to optimize the upper bound, $\tilde{J}( \boldsymbol{\Gamma}, \boldsymbol{\Theta}_{1:N_C})$, including the expectation step (E-step) and the maximization step (M-step). In the E-step, we calculate the conditional distribution,  $\wp_{C \vert \mathbf{X}^*} (C 
= i \vert \mathbf{x}_n,  \boldsymbol{\Theta}_{1:N_C}^{\ell} )$, for every sample, $\mathbf{x}_n$, given the parameter, $\boldsymbol{\Theta}_{1:N_C}^{\ell}$, obtained at the $\ell$ iteration, such that
\begin{align}
	\gamma_{n,i}^{\ell} & =  \wp_{C \vert \mathbf{X}^*} (C 
	= i \vert \mathbf{x}_n,  \boldsymbol{\Theta}_{1:N_C}^{\ell} )  \nonumber\\
	&= \frac{\omega_i^{\ell} f_{\mathbf{X}^*, i} (\mathbf{x}_n \vert \boldsymbol{\mu}_i^{\ell}, \boldsymbol{\Sigma}_i^{\ell}) }{\sum_{i=1}^{N_C} \omega_i^{\ell} f_{\mathbf{X}^*, i} (\mathbf{x}_n \vert \boldsymbol{\mu}_i^{\ell}, \boldsymbol{\Sigma}_i^{\ell})}, \text{ for } i = 1, \ldots, N_C
	\label{eq: gamma_n_i}
\end{align} 

In the M-step, fixing $\boldsymbol{\Gamma}^{\ell} =  \begin{bmatrix}
\gamma_{n,i}^{\ell} \end{bmatrix}_{n=1, i = 1}^{N, N_C}$ obtained in the E-step, we minimize $\tilde{J}( \boldsymbol{\Gamma}^{\ell}, \boldsymbol{\Theta}_{1:N_C})$ w.r.t. the parameters, $\boldsymbol{\Theta}_{1:N_C}$,  for the $(\ell + 1)$th iteration, such that
\begin{equation}
	\boldsymbol{\Theta}_{1:N_C}^{\ell + 1} = \underset{\boldsymbol{\Theta}_{1:N_C}}{\arg\min} \tilde{J}(\boldsymbol{\Gamma}^{\ell}, \boldsymbol{\Theta}_{1:N_C}) 
\end{equation}
Considering that $\sum_{i=1}^{N_C} \omega_i = 1$ and $\omega_i > 0$ for all $i$, $\boldsymbol{\omega}^{\ell + 1} = \begin{bmatrix}
	\omega_1^{\ell+1} & \ldots & \omega_{N_C}^{\ell+1}
\end{bmatrix}$ can be updated by
\begin{equation}
	\label{eq: cal_omega_k}
	\omega_i^{\ell+1} = \frac{\sum_{n=1}^N \gamma_{n,i}^{\ell}}{\sum_{n=1}^N \sum_{i=1}^{N_C}  \gamma_{n,i}^{\ell}} = \frac{\sum_{n=1}^N \gamma_{n,i}^{\ell}}{N} 
	\text{ for } i = 1,\ldots, N_C  
\end{equation}

Furthermore, we use the gradient-decent method to obtain the parameters, $\boldsymbol{\Theta}_i^{\ell + 1} = (\boldsymbol{\mu}_i^{\ell + 1}, \boldsymbol{\Sigma}_i^{\ell + 1})$ by calculating the partial derivatives of $\tilde{J}(\boldsymbol{\Gamma}^{\ell}, \boldsymbol{\Theta}_{1:N_C})$ w.r.t. $\boldsymbol{\mu}_i$ and $\boldsymbol{\Sigma}_i$, recursively, at the $\ell$th learning iteration. Considering the constraint that $\boldsymbol{\Sigma}_i^{\ell + 1}$ must be a  positive definite matrix, we update $\boldsymbol{\Sigma}_i^{\ell + 1}$ using the partial derivative w.r.t. the matrix $\mathbf{L}_i$ instead of $\boldsymbol{\Sigma}_i$, where 
\begin{equation}
	\boldsymbol{\Sigma}_i^{-1} = \mathbf{L}_i\mathbf{L}_i^T
	\label{eq: cholesky_inverse_Covariance}
\end{equation} 
and the superscript ``$T$" indicates the transpose operation. Specifically, the derivatives of $\tilde{J}(\boldsymbol{\Gamma}^{\ell}, \boldsymbol{\Theta}_{1:K})$ w.r.t. $\boldsymbol{\mu}_i$ and $\mathbf{L}_i$ can be expressed by
\begin{equation}
	\frac{\partial \tilde{J} }{\partial \boldsymbol{\mu}_i} 
	= \omega_i^{\ell+1} N \boldsymbol{\Sigma}_i^{-1}  \bigg[  \frac{\mathbb{E}_{\mathbf{X} \sim \phi_{\mathbf{X}, i }}[  \mathbf{X} Q_Y(\mathbf{X})]}{\mathbb{E}_{\mathbf{X}\sim \phi_{\mathbf{X}, i}}[Q_Y(\mathbf{X})]} - \hat{\boldsymbol{\mu}}_i \bigg]
	\label{eq: partial_tilde_J_mu}
\end{equation}
\begin{align}
	\frac{\partial \tilde{J} }{\partial \mathbf{L}_i} 	=& -\omega_i^{\ell+1} N   \bigg\{ \frac{\mathbb{E}_{\mathbf{X}\sim \phi_{\mathbf{X}, i}}[(\mathbf{X} 
		- \boldsymbol{\mu}_i) (\mathbf{X} - \boldsymbol{\mu}_i)^T Q_Y(\mathbf{X})]}{\mathbb{E}_{\mathbf{X}\sim \phi_{\mathbf{X}, i}}[Q_Y(\mathbf{X})]}   \nonumber\\
	&-  \hat{\boldsymbol{\Sigma}}_i(\boldsymbol{\mu}_i^{\ell}) \bigg\} \mathbf{L}_i
	\label{eq: partial_tilde_J_L}
\end{align}
where  
\begin{align}
	\hat{\boldsymbol{\mu}}_i &=  \frac{ \sum_{i=1}^N\gamma_{n,i}^{\ell} \mathbf{x}_n}{\sum_{n=1}^N \gamma_{n,i}^{\ell}}  \\
	\hat{\boldsymbol{\Sigma}}_i(\boldsymbol{\mu}_i^{\ell}) &=   \frac{ \sum_{n=1}^N \gamma_{n,i}^{\ell}  (\mathbf{x}_n - \boldsymbol{\mu}_i^{\ell}) (\mathbf{x}_n - \boldsymbol{\mu}_i^{\ell})^T}{\sum_{n=1}^N \gamma_{n,i}^{\ell}} 
\end{align}
Because the skewing function, $Q_Y$, is an arbitrary function determined by the layout of obstacles, there are no closed forms for calculating the expectation and $\gamma_{n,i}$ terms, which all involve integrals. Therefore, these terms must be approximated to implement the proposed learning algorithm. This paper uses $M$ auxiliary samples to approximate the integral calculations. For example, the integral of an arbitrary function $g(\mathbf{x})$ over $\mathcal{W}$ can be approximated by
	$\int_{\mathcal{W}} g(\mathbf{x}) d\mathbf{x} \approx \sum_{j=1}^M g(\boldsymbol{\xi}_j) \Delta x \Delta y$,   
where $\mathcal{D}_A = \{ \boldsymbol{\xi}_j\}_{j=1}^M$ denotes the set of $M$ grid points evenly deployed on the 2-D workspace with the spatial resolution of $\Delta x$ and $\Delta y$.  

Finally, we can update the parameters to obtain $\boldsymbol{\Theta}_i^{\ell+1} = (\boldsymbol{\mu}_i^{\ell + 1}, \boldsymbol{\Sigma}_i^{\ell + 1})$ for $i = 1, \ldots, N_C$ by
\begin{align}
	\boldsymbol{\mu}_i^{\ell + 1} &= \boldsymbol{\mu}_i^{\ell} - \lambda_{\boldsymbol{\mu}} \frac{\partial \tilde{J}(\boldsymbol{\Theta}^{\ell})}{\partial \boldsymbol{\mu}_i} \label{eq: update_mu}\\
	\mathbf{L}_i^{\ell + 1} &= \mathbf{L}_i^{\ell} - \lambda_{\mathbf{L}} \frac{\partial \tilde{J}(\boldsymbol{\Theta}^{\ell})}{\partial \mathbf{L}_i} \\
	\boldsymbol{\Sigma}_i^{\ell + 1} &= \bigg( \mathbf{L}_i^{\ell + 1} {\mathbf{L}_i^{\ell + 1}}^T \bigg)^{-1} \label{eq: update_Sigma}
\end{align}
where $\lambda_{\boldsymbol{\mu}}, \lambda_{\mathbf{L}} \in \mathbb{R}^+$ are the user-defined learning rates. The proposed parameter learning algorithm is summarized in \textbf{Algorithm} \ref{alg: param_learning_BRF_SNMM}, where double loops ($\ell_1$ and $\ell_2$) are applied to update $\boldsymbol{\omega}$ and $\boldsymbol{\Theta}_i$ asynchronously. It is worth mentioning that the proposed algorithm can also be utilized to learn the parameters for the BRF-SN distribution since the BRF-SN can be treated as a special BRF-SNMM with $N_C = 1$ and $\boldsymbol{\omega} = 1$. 
\begin{algorithm}
	\caption{Parameter Learning for BRF-SNMM} 
	\label{alg: param_learning_BRF_SNMM}
	\begin{algorithmic}[1]
		\REQUIRE  $\qquad$ \\
		Sample data set: $\mathcal{D} = \{\mathbf{x}_n\}_n^N$ \\
		Auxiliary sample data set: $\mathcal{D}_A = \{\boldsymbol{\xi}_j\}_{j=1}^M$ \\
		Skewing function: $Q(\mathbf{x})$ defined on $\mathcal{W} \in \mathbb{R}^{n_x}$\\
		Set the number of mixture components: $N_C$ \\
		Initialize the iteration indices: $\ell_1 \leftarrow 0$ and $\ell_2 \leftarrow 0$\\
		Initialized parameters: $\boldsymbol{\Theta}_{1:N_C}^{\ell_1} = \{ (\omega_i^{\ell_1}, \boldsymbol{\Theta}_i^{\ell_2} )\}_{i=1}^{N_C}$ \\
		Approximate the upper bound of the cost function: $\tilde{J}( \boldsymbol{\Gamma}, \boldsymbol{\Theta}_{1:N_C}^{\ell_1})$ according to (\ref{eq: costFunc_paramLearning})\\
		Set the termination parameters: $L_1$ and $L_2$ \\
		Set the learning rates: $\lambda_{\boldsymbol{\mu}}$ and $\lambda_{\mathbf{L}}$
		\WHILE{($\ell_1 < L_1$)  }	
		\STATE Approximate the matrix $\boldsymbol{\Gamma}^{\ell_1} = \begin{bmatrix}
			\gamma_{n,i}^{\ell_1}
		\end{bmatrix}_{n=1, i=1}^{N, N_C}$ according to (\ref{eq: gamma_n_i}) given  $\boldsymbol{\Theta}_{1:N_C}^{\ell_1}$ and $\mathcal{D}_A$
		\STATE Calculate component weights, $\boldsymbol{\omega}^{\ell_1 + 1}$, according to (\ref{eq: cal_omega_k})
		\STATE Update parameters: $\boldsymbol{\Theta}_{1:N_C}^{\ell_1+1} = \{ (\omega_i^{\ell_1 + 1}, \boldsymbol{\Theta}_i^{\ell_2} )\}_{i=1}^{N_C}$
		\FOR{$i = 1$ to $N_C$}
		\STATE Reset the iteration index: $\ell_2 \leftarrow 0$\\
		\WHILE{($\ell_2 < L_2$) }
		\STATE Obtain $\mathbf{L}_i^{\ell_2}$ based on $\boldsymbol{\Sigma}_i^{\ell_2}$ according (\ref{eq: cholesky_inverse_Covariance})
		\STATE Approximate $\frac{\partial \tilde{J}(\boldsymbol{\Gamma}^{\ell_1}, \boldsymbol{\Theta}^{\ell_2}_{1:N_C}) }{\partial \boldsymbol{\mu}_i^{\ell_2}}$ and  $\frac{\partial \tilde{J}(\boldsymbol{\Gamma}^{\ell_1}, \boldsymbol{\Theta}^{\ell_2}_{1:N_C}) }{\partial \mathbf{L}_i^{\ell_2}}$, based on $\mathcal{D}_A^{\ell}$ according to (\ref{eq: partial_tilde_J_mu}) and (\ref{eq: partial_tilde_J_L})
		
		\STATE Update $\boldsymbol{\Theta}_i^{\ell_2+1} = (\boldsymbol{\mu}_i^{\ell_2 + 1}, \boldsymbol{\Sigma}_i^{\ell_2 + 1})$,  according to (\ref{eq: update_mu}) - (\ref{eq: update_Sigma}) 
		\STATE Update the iteration index: $\ell_2 \leftarrow \ell_2 + 1$
		\STATE Update parameters: $\boldsymbol{\Theta}_i^{\ell_2} \leftarrow \boldsymbol{\Theta}_i^{\ell_2 + 1}$
		\ENDWHILE
		\ENDFOR
		\STATE Update the iteration index: $\ell_1 \leftarrow \ell_1 + 1$
		\ENDWHILE
	\end{algorithmic}
\end{algorithm}

\section{PATH-PLANNING BASED ON SKEW-NORMAL MIXTURE MODEL}
\label{sec: PathPlanning}
In this section, we apply the proposed BRF-SNMM to solve the path-planning problems for VLSR systems in a known and time-invariant 2-D obstacle-deployed environment. The VLSR systems comprise $N$ cooperative agents, whose microscopic dynamics are 
modeled by the following stochastic differential equation (SDE), 
\begin{align}
	\dot{\mathbf{x}}_n(t) &= \mathbf{f}[\mathbf{x}_n(t), \mathbf{u}_n(t), t] \label{eq:individual_SDE}\\
	\mathbf{x}_n(t_0) &= \mathbf{x}_{n,0}, n = 1,\ldots, N \label{eq:individual_Init}
\end{align}
where $\mathbf{x}_n(t) \in \mathcal{W}$ and $\mathbf{u}_n(t) \in \mathcal{U}$ are the $n$th agent's microscopic state and control at time $t$, respectively, and $\mathbf{x}_{n,0}$ is the $n$th agent's initial state. It is assumed that the agents' states are all fully observable.

The agents’ PDF represents the macroscopic state of the VLSR systems. On the macroscopic scale, the path-planning task of the VLSR systems can be formulated as the distribution path-planning problem, where a trajectory of agents' PDFs in the workspace over a time interval $[t_0, t_f]$ is obtained to guide the agents' microscopic controls. Specifically, the agents' PDF at time $t \in [t_0, t_f]$, denoted by $\wp(\mathbf{x}, t)$, is modeled by a time-varying BRF-SNMM, such that 
\begin{equation}
\wp(\mathbf{x}, t) \triangleq \wp_{\mathbf{X}^*}\big(\mathbf{x} \vert \boldsymbol{\Theta}_{1:N_C}(t) \big) = \sum_{i=1}^{N_C} \omega_i f_{\mathbf{X}^*, i} \big(\boldsymbol{\Theta}_i(t) \big) 
\label{eq:SNMM_agent_PDF}
\end{equation}
where $\boldsymbol{\Theta}_i(t) = \big(\boldsymbol{\mu}_i(t), \boldsymbol{\Sigma}_i(t)\big)$ and $\boldsymbol{\Theta}_{1:N_C}(t) = \{ (\omega_i, \boldsymbol{\Theta}_i(t) )\}$. 
Therefore, given the skewing function $Q_Y$ at time $t$, the time-varying BRF-SNMM $\wp(\mathbf{x}, t)$ is specified by the parameters, $N_C$ and $\boldsymbol{\Theta}_{1:N_C}(t) $. 

For simplicity, we have three assumptions for this path-planning problem: 
\begin{itemize}
	\item \textit{A.1} The component number $N_C$ and the component weights $\boldsymbol{\omega} = \begin{bmatrix}
		\omega_1 & \ldots & \omega_{N_C}
	\end{bmatrix}$ in (\ref{eq:SNMM_agent_PDF}) are all fixed and known constants. 
	\item \textit{A.2} The desired agents' distribution is a BRF-SN such that $\wp_f = f_{\mathbf{X}^*}(\mathbf{x} \vert \boldsymbol{\Theta}_f) $, specified by the parameter set, $\boldsymbol{\Theta}_f = (\boldsymbol{\mu}_f, \boldsymbol{\Sigma}_f)$. 
	\item \textit{A.3} The obstacles deployed in the workspace can all be represented geometrically by convex polygons, circles, or ovals. 
\end{itemize}
%
Here, we introduce Assumption \textit{A.3} to alleviate the difficulty of the path-planning problem by preventing robots from becoming trapped in local minima.

In this section, we focus on the macroscopic scale path-planning task and propose two approaches to generate the trajectories of agents' BRF-SNMM-based distributions from  $\wp(\mathbf{x}, t_0)$ to $\wp_f$. The microscopic control is only described briefly in this paper, and the readers are referred to our previous papers for more details \cite{zhu2021adaptive, zhu2019scalable}.

\subsection{Distribution Path-Planning based on Displacement Interpolation}

Because the parameter set $\boldsymbol{\Theta}_{1:N_C}(t)$ in (\ref{eq:SNMM_agent_PDF}) and the parameter set $\boldsymbol{\Theta}_f$ in Assumption A.2 can also specify a time-varying GMM, $\varphi(\mathbf{x}, t) \triangleq \sum_{i=1}^{N_C} \omega_i \phi_{\mathbf{X},i}\big(\mathbf{x} \vert \boldsymbol{\Theta}_i(t)\big)$, and a desired normal distribution, $\phi_f = \phi_{\mathbf{X}}(\mathbf{x} \vert \boldsymbol{\Theta}_f)$, respectively. We can provide a two-step approach to the path-planning problem. We can, first, obtain a trajectory of the time-varying GMM from the start GMM $\varphi(\mathbf{x}, t_0)$ to the desired normal distribution $\phi_f$ in the obstacle-free workspace,  then form the time-varying BRF-SNMM trajectory based on the obtained time-varying parameter set, $\boldsymbol{\Theta}_{1:N_C}(t)$, according to (\ref{eq:SNMM_agent_PDF}).


In this paper, we apply the displacement interpolation (DI) between $\varphi(\mathbf{x}, t_0)$ and  $\phi_f$ to specify the time-varying parameter set, $\boldsymbol{\Theta}_i(t) = \big(\boldsymbol{\mu}_i(t), \boldsymbol{\Sigma}_i(t)\big)$ for $i=1,\ldots, N_C$ \cite{chen2018optimal}, such that
\begin{align}
	\boldsymbol{\mu}_i(t) &= \frac{t_f - t_0 - t}{t_f - t_0} \boldsymbol{\mu}_i(t_0) + \frac{t}{t_f - t_0} \boldsymbol{\mu}_f \\
	\boldsymbol{\Sigma}_i(t) &= \boldsymbol{\Sigma}_i(t_0)^{-1/2} \bigg\{ \frac{t_f - t_0 - t}{t_f - t_0} \boldsymbol{\Sigma}_i(t_0)  +  \frac{t}{t_f - t_0} \nonumber\\
	& \cdot \bigg[\boldsymbol{\Sigma}_i(t_0)^{1/2}  \boldsymbol{\Sigma}_f  \boldsymbol{\Sigma}_i(t_0)^{1/2}  \bigg]^{1/2} \bigg\}^2 \boldsymbol{\Sigma}_i(t_0)^{-1/2} 
\end{align} 
This approach is referred to as the SNMM-DI approach. Although the obtained GMM trajectory, $\varphi(\mathbf{x}, t \vert \boldsymbol{\Theta}_{1:N_C}(t))$, is a geodesic path, which provides the shortest $\ell-2$ Wasserstein distance \cite{chen2018optimal}, it is not guaranteed that the obtained BRF-SNN trajectory can give the shortest $\ell-2$ Wasserstein distance in the obstacle-deployed environment since the time-varying parameter set  is obtained without considering the obstacles.

\subsection{Distribution Path-Planning based on Artificial Potential Field}
Considering that the environment/obstacle information is embedded in the BRF-SN and BRF-SNMM distributions, the Artificial Potential Field (APF) method can be directly applied to solve the path planning problem without special consideration of the influence of obstacles. In this subsection, two APFs, as well as their derivatives, are developed. The parameter set,  $\boldsymbol{\Theta}_i(t) = \big(\boldsymbol{\mu}_i(t), \boldsymbol{\Sigma}_i(t)\big)$ for $i=1,\ldots, N_C$, is approximated to generate the BRF-SNMM trajectory. 

First, to reduce the difference between the current agents' distribution and the desired BRF-SN distribution, the following potential based on the square of the functional  $\ell - 2$ norm is applied, 
\begin{equation}
	U_{SN}(t) \triangleq \frac{1}{2} \ln \int_{\mathcal{W}} \bigg[\wp_{\mathbf{X}^*} (\mathbf{x}, t) - f_{\mathbf{X}^*} (\mathbf{x} \vert \boldsymbol{\Theta}_f ) \bigg]^2 d\mathbf{x}
	\label{eq: log_potential_ED_SNMM}
\end{equation} 
Similarly to \textbf{Algorithm} \ref{alg: param_learning_BRF_SNMM}, the gradient-decent method is applied to generate the trajectory of agents' distributions by minimizing the potential w.r.t. the time-varying parameters. Specifically, the partial derivatives w.r.t. the parameters are derived and approximated based on the auxiliary grid samples, $\mathcal{D}_A$. Here, the logarithm is utilized in (\ref{eq: log_potential_ED_SNMM}) to cancel the spatial resolutions, $\Delta x$ and $\Delta y$, in the approximations of the partial derivatives.

Moreover, because $\wp_{\mathbf{X}^*}(\mathbf{x}, t)$ defined in (\ref{eq:SNMM_agent_PDF}) can be treated as a weighted linear combination of its BRF-SN components, $\{f_{\mathbf{X}^*_i} (\cdot \vert \boldsymbol{\Theta}_i(t) )\}_{i=1}^{N_C}$, associated with the component weights, $\boldsymbol{\omega}$. We propose the other potential based on the Cauchy-Schwarz (CS) divergence between the BRF-SN distributions, such that 
\begin{equation}
	U_{CS}(t) \triangleq \sum_{i=1}^{N_C} \omega_i D_{CS}\bigg(f_{\mathbf{X}^*, i} (\cdot \vert \boldsymbol{\Theta}_i(t) )  \bigr\Vert f_{\mathbf{X}^*} (\cdot \vert \boldsymbol{\Theta}_f )  \bigg) 
\end{equation}
where the term $D_{CS}(f_1  \bigr\Vert f_2  )$ denotes the CS-divergence between two PDFs, $f_1$ and $f_2$ \cite{kampa2011closed}, defined by
\begin{align}
	D_{CS}(f_1  \bigr\Vert f_2  ) \triangleq -\ln \frac{\int f_1(\mathbf{x})f_2(\mathbf{x}) d \mathbf{x})}{\sqrt{\int f_1^2(\mathbf{x}) d\mathbf{x}\int f_2^2(\mathbf{x}) d\mathbf{x}}} 
\end{align}
It is noteworthy that due to the ratio-form definition of the CS-divergence, the spatial resolutions,  $\Delta x$ and $\Delta y$, are also canceled in the approximations of the partial derivatives w.r.t. the parameters $\boldsymbol{\Theta}_i(t)$ for $i=1,\ldots,N_C$.  

Furthermore, we notice that the potential, $U_{SN}(t)$, provides a more significant force when the agents' distribution is close to the desired distribution due to the logarithm function. In contrast, the potential,   $U_{CS}(t)$, can contribute a more significant force at the initial stage of the distribution path planning problem. Therefore, we combine these two potentials to create a novel potential to generate the trajectory of agents' distributions, such that
\begin{equation}
	U(t) \triangleq  \gamma_{SN} U_{SN}(t) + \gamma_{CS} U_{CS}(t)
	\label{eq: potential}
\end{equation} 
where $\gamma_{SN}, \gamma_{CS} \in \mathbb{R}^+$ are user-defined parameters. 

Finally, the partial derivatives of $U(t)$ defined in (\ref{eq: potential}) w.r.t. the parameters $\boldsymbol{\Theta}_{1:N_C}(t)$ are applied to obtain the trajectory of agents PDFs similar to (\ref{eq: update_mu}) - (\ref{eq: update_Sigma}). The implementation details are omitted due to the limited space. This approach is referred to as the SNMM-APF approach in this paper.  

\subsection{Microscopic Control for Individual Agent}
Given the optimal agents' distribution, $\wp(\mathbf{x}, t)$, obtained by minimizing the potential, $U(t)$, the corresponding microscopic control, denoted by $\mathbf{u}_n(t)$ for $n = 1, \ldots, N$ in (\ref{eq:individual_SDE}), can be generated by minimizing the following weighted-linear-combination potential under the dynamic constraints given in (\ref{eq:individual_SDE}) and (\ref{eq:individual_Init}), 
\begin{equation}
	U_{MC} (t) = \gamma_{Att} U_{Att}(t) + \gamma_{CA} U_{CA}(t)
\end{equation}  
where $\gamma_{Att}, \gamma_{CA} \in \mathbb{R}^{+}$ are user-define parameters for the attractive and the collision avoidance potentials, $U_{Att}$ and $U_{CA}$, respectively. The attractive potential is designed to ``push" the agents toward the optimal agents' distribution, such that
\begin{equation}
	U_{Att}(t)\triangleq \int_{\mathcal{W}} \bigg[ \hat{\wp}(\mathbf{x}, t)- \wp(\mathbf{x}, t)\bigg]^2 d\mathbf{x}
\end{equation}
where $\hat{\wp}(\mathbf{x}, t)$ is the approximate of the agents' PDF based on agents' positions by the kernel density estimation (KDE) method. Then, the collision avoidance potential is applied to generate the repulsive force against the obstacles and other agents, which is defined by
\begin{equation}
	U_{CA}(t) = \sum_{n=1}^N \bigg[ \frac{1}{\rho_n(t)} - \frac{1}{\rho_0}\bigg]^2 \boldsymbol{1}\big(\rho_n(t), \rho_0\big) 
\end{equation}   
where $\rho_n$ indicates the shortest distance between the $n$th agent and the obstacles or  other agents, $\rho_0$ is a tolerance threshold, and $\boldsymbol{1}(\rho_n, \rho_0)$ is an indicator function that equals one if $\rho_n \leq \rho_0$, and equals zero otherwise.   

\section{EXPERIMENTS AND RESULTS}
 
Two experiments are provided in this section to demonstrate the effectiveness of the proposed parameter learning algorithm in Section \ref{subsec: paramLearning_SNMM} and the BRF-SNMM-based path-planning approaches in Section \ref{sec: PathPlanning}, respectively.

\subsection{Comparison between SNMM and GMM}
\label{sec: Exp-A}
\begin{figure}
	\centering
	\includegraphics[width=\figsize\textwidth]{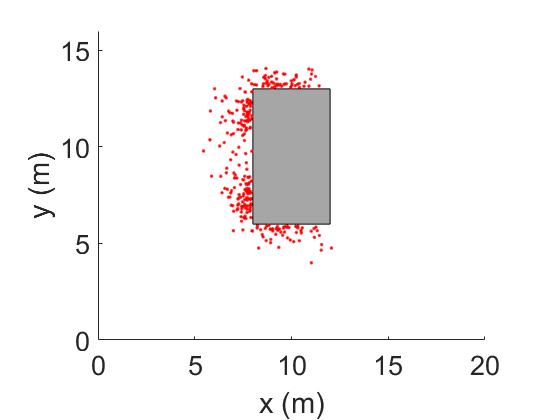}  
	\includegraphics[width=\figsize\textwidth]{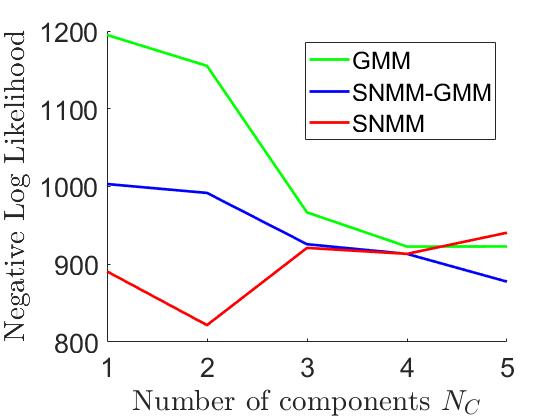}
	\caption{ Left: Samples ($N = 300$) generated according to the underlying 2-component  BRF-SNMM distribution are deployed with the obstacle in the workspace. The red points indicate the samples, and the gray rectangle indicates the obstacle. Right: Comparison of parameter learning performances. The NLLs are generated by different approaches for different values of $N_C$.}
	\label{fig:Simulation-I}
\end{figure}

To evaluate the learning ability of  \textbf{Algorithm} \ref{alg: param_learning_BRF_SNMM}, $N = 300$ samples, $\mathcal{D} = \{\mathbf{x}_n\}_{n=1}^N$, are generated according to a  2-component BRF-SNMM distribution, $\wp_{\mathbf{X}^*}(\mathbf{x} \vert \boldsymbol{\Theta}_{1:N_C})$ with $N_C = 2$, and deployed with the obstacle, as shown in Fig. \ref{fig:Simulation-I} (Left). For simplicity, a binary skewing function, $Q_Y(\mathbf{x}) \in \{0, 1\}$, is applied to indicate if the obstacle in Fig. \ref{fig:Simulation-I} (Left)  occupies the location. The parameters of the BRF-SNMM, $\boldsymbol{\Theta}_{1:N_C} = \{(\omega_i, \boldsymbol{\Theta}_i)\}_{i=1}^{2}$ are set as follows, $\omega_1 = \omega_2 = 0.5$, $\boldsymbol{\mu}_1 = \begin{bmatrix}
	9 & 12
\end{bmatrix}$, $\boldsymbol{\Sigma}_1 = \begin{bmatrix}
1 & 0.3\\ 0.3 & 0.7
\end{bmatrix}$,  $\boldsymbol{\mu}_2 = \begin{bmatrix}
9 & 7
\end{bmatrix}$, and $\boldsymbol{\Sigma}_2 = \begin{bmatrix}
1 & -0.3\\ -0.3 & 0.7
\end{bmatrix}$.  

In this experiment, only the sample data set, $\mathcal{D}$, is available, and  the parameters, $N_C$, and $\boldsymbol{\Theta}_{1:N_C} = \{ (\omega_i, \boldsymbol{\Theta}_i )\}_{i=1}^{N_C}$, all need to be estimated.  We try five component numbers, such that $N_C = 1, 2, 3, 4, 5$, to run \textbf{Algorithm} \ref{alg: param_learning_BRF_SNMM}.

The learning performance of \textbf{Algorithm} \ref{alg: param_learning_BRF_SNMM} is compared with the other two approaches. 1) \textbf{GMM} approach: Assume that the samples are generated by an underlying GMM and the GMM’s parameters are estimated from $\mathcal{D}$, 2) \textbf{SNMM-GMM} approach: Assume that the samples are generated by an underlying BRF-SNMM and the parameters $\boldsymbol{\Theta}_{1:N_C}$ are estimated using the GMM parameter learning algorithm. For a fair comparison, in \textbf{Algorithm} \ref{alg: param_learning_BRF_SNMM}, the SNMM is also initialized by the GMM's parameter learning algorithm, specially, the iterative Expectation-Maximization (EM) algorithm \cite{peel2000GMM}. 

The NLLs of the samples generated based on different approaches are plotted in Fig. \ref{fig:Simulation-I} (Right). It can be seen that only our algorithm can provide the smallest NLL at $N_C = 2$. It means that the statistical model learned by our proposed Algorithm 1 can describe the samples using only two components, as equal to the component number of the ground-truth SNMM distribution. In contrast, the SNMM-GMM and GMM approaches request four or more components to describe the samples.

\subsection{Path-Planning for VLSR System}
\label{sec: Exp-B}

\begin{figure}
	\centering
	\includegraphics[width=\figsize \textwidth]{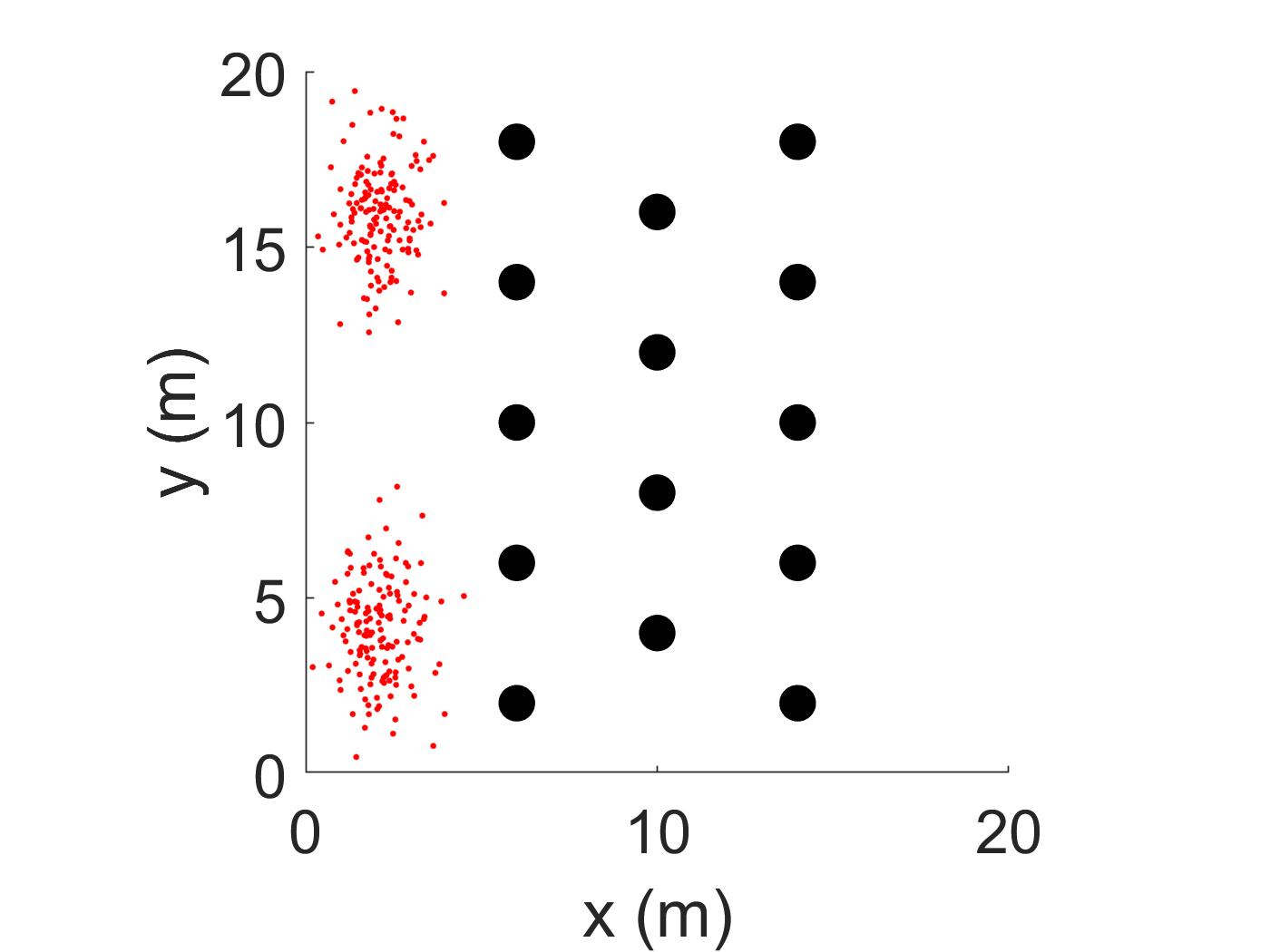}  
	\includegraphics[width=\figsize \textwidth]{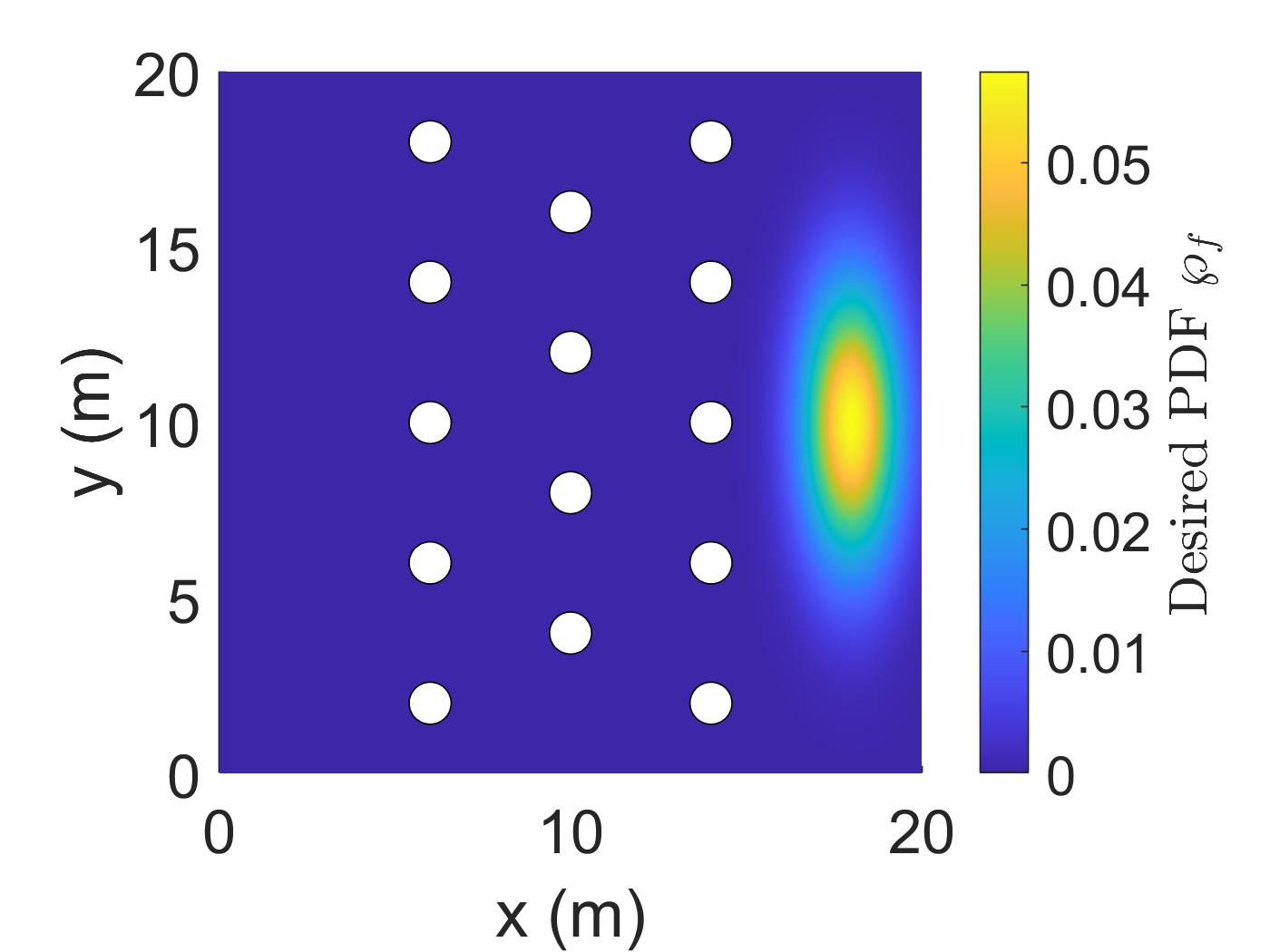} 
	\caption{The path-planning problem for VLSR systems in the simple artificial forest
		environment (Forest-I simulation). Left: The red points indicate the initially deployed agents, and the black circles represent the “trees” in the forest. Right: The desired agents’ PDF is shown, where the white areas are occupied by “trees”.   }
	\label{fig:artifical_forest}
\end{figure}

To evaluate the proposed path-planning algorithms, first, we create a simple artificial forest environment (Forest-I) shown in Fig. \ref{fig:artifical_forest} (left), where 14 black circles indicate the trees in the artificial forest. A similar artificial forest environment has been recently applied and reported in \cite{soria2021distributed}. However, unlike \cite{soria2021distributed} where only 16 drones are simulated, a VLSR system comprised of $N=300$ agents is tested in this experiment. Given the agents' initial positions in the artificial forest, the red points indicated by the red points in Fig. \ref{fig:artifical_forest} (left), the goal of the path-planning task is to guide all of the agents to traverse the artificial forest and avoid collisions with these "trees" and achieve the desired distribution, $\wp_f$, shown in Fig. \ref{fig:artifical_forest} (right). The white circles shown in Fig. \ref{fig:artifical_forest} (right) indicate the areas where $Q_Y(\mathbf{x})=0$, which means that agents are not allowed to go through these areas because they are occupied by "trees." 

\begin{figure}[ht]
	\centering
	\includegraphics[width=\figsize\textwidth]{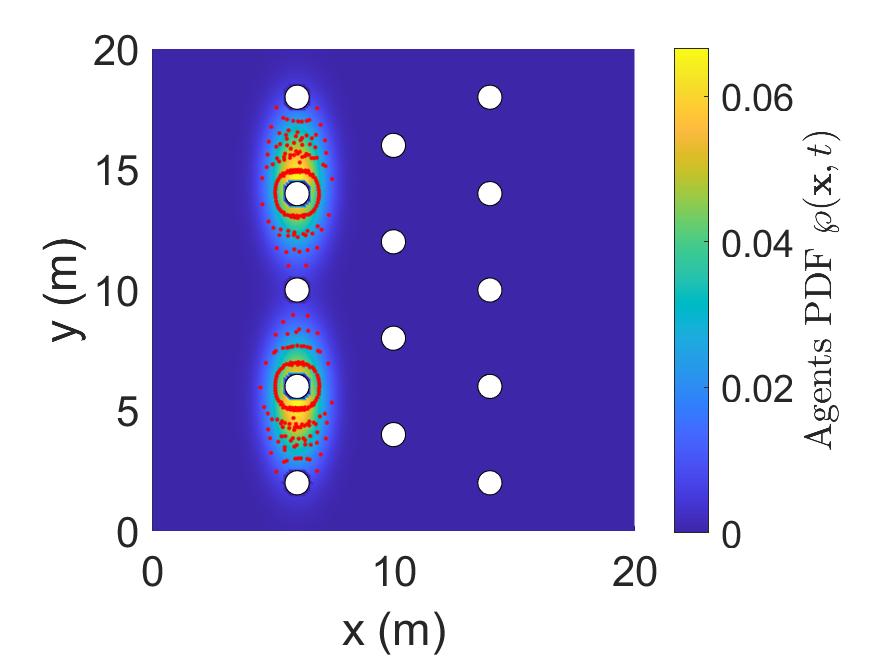} 
	\includegraphics[width=\figsize\textwidth]{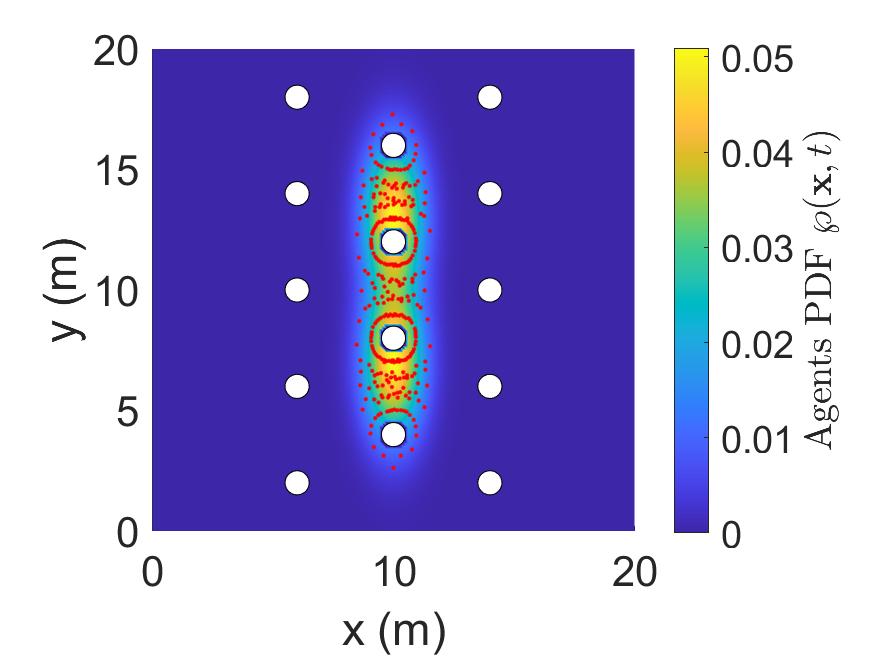} 
	\includegraphics[width=\figsize\textwidth]{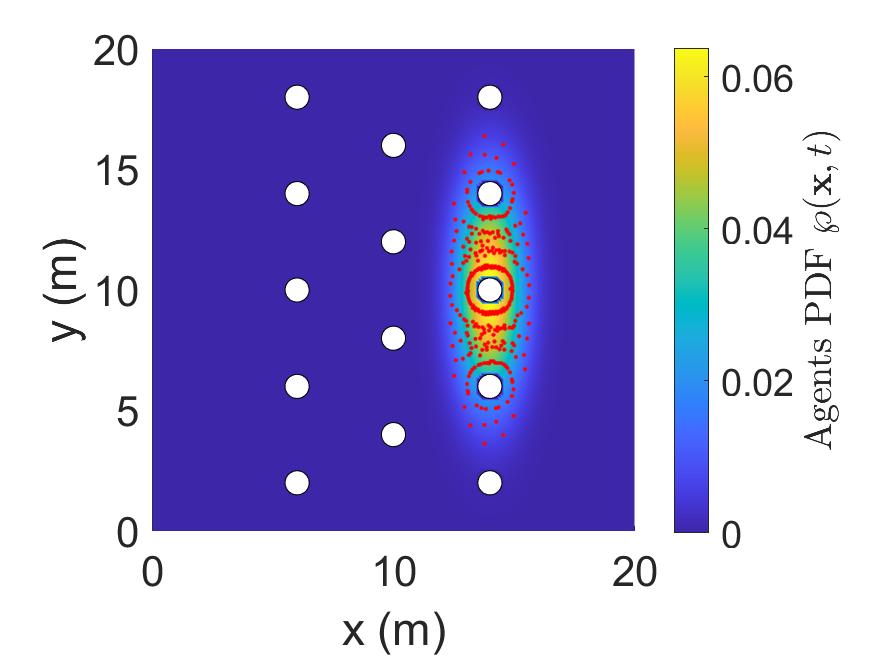} 
	\includegraphics[width=\figsize\textwidth]{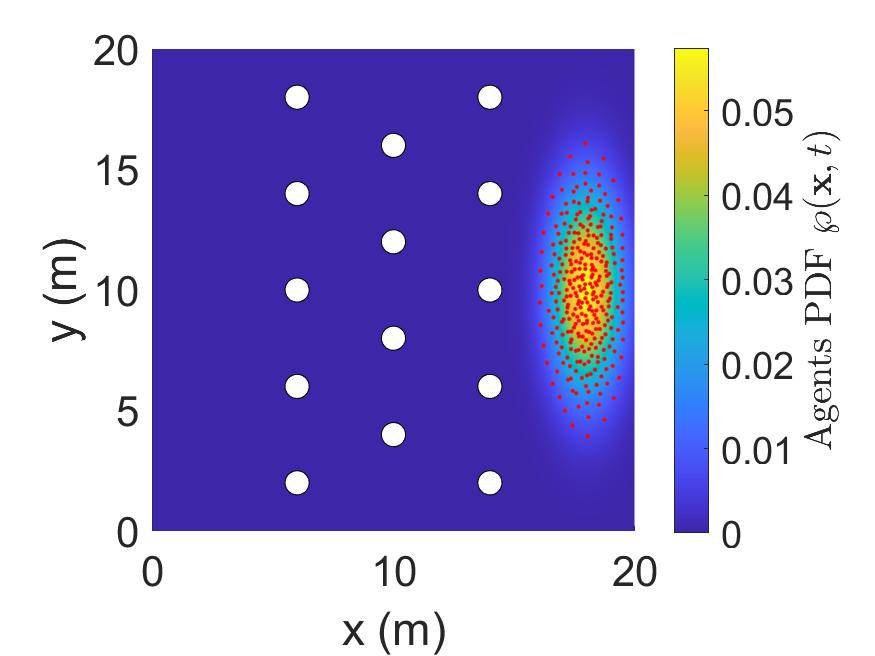} 
	\caption{Snapshots of the trajectory of agents and corresponding PDFs generated by the SNMM-DI approach in the Forest-I simulation. }
	\label{fig:Snapshots_interpolation}
\end{figure}

\begin{figure}[ht]
	\centering
	\includegraphics[width=\figsize\textwidth]{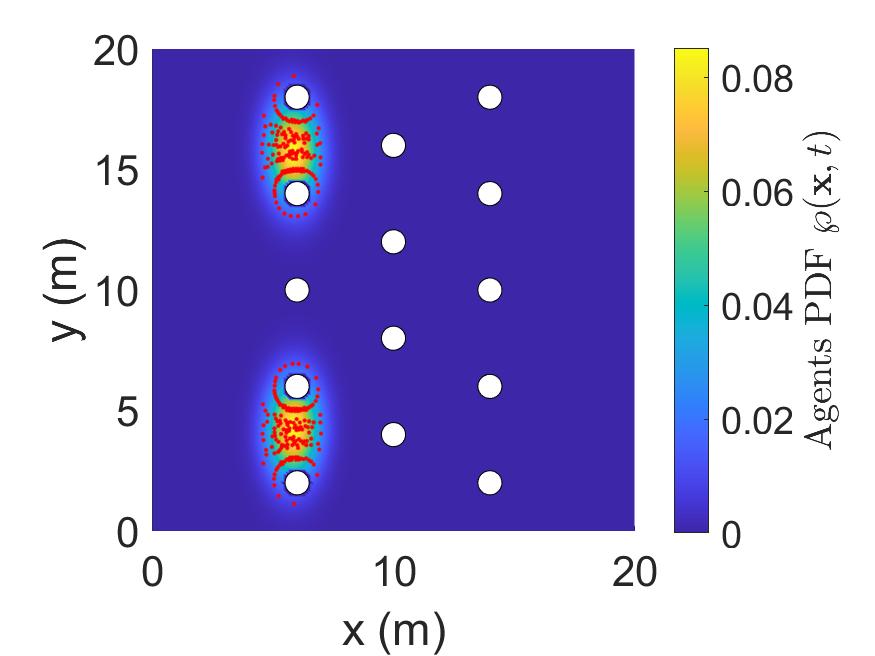} 
	\includegraphics[width=\figsize\textwidth]{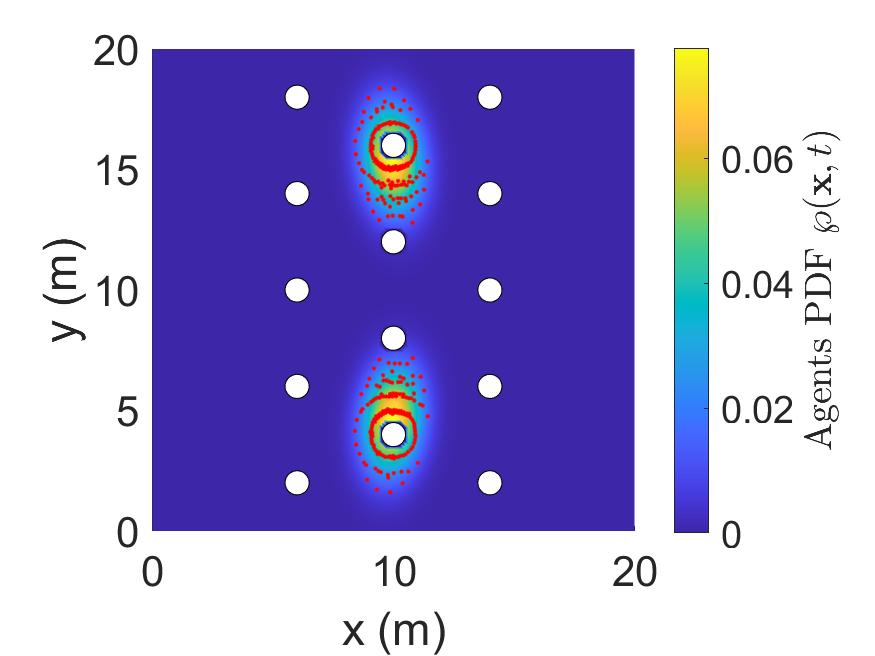} 
	\includegraphics[width=\figsize\textwidth]{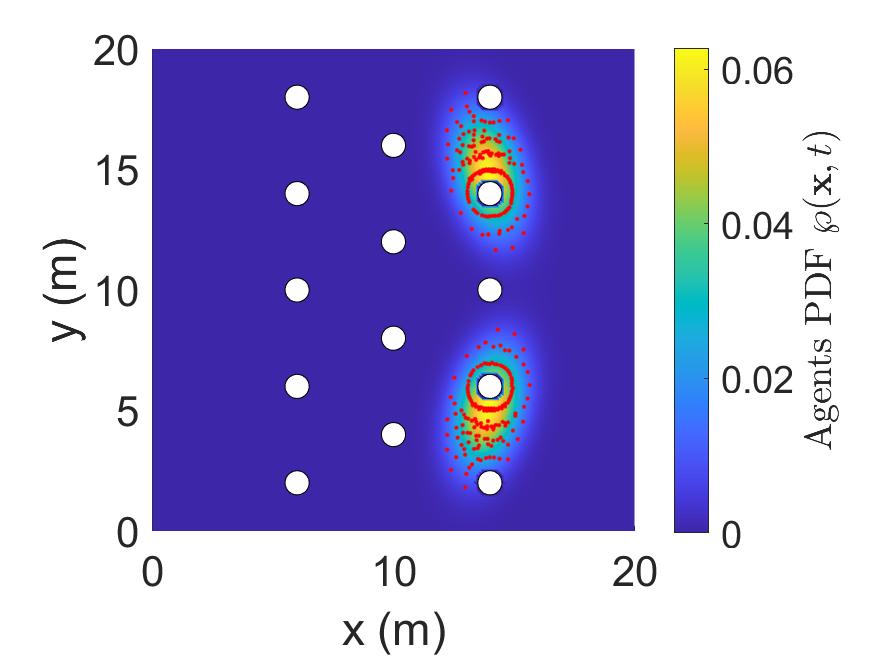} 
	\includegraphics[width=\figsize\textwidth]{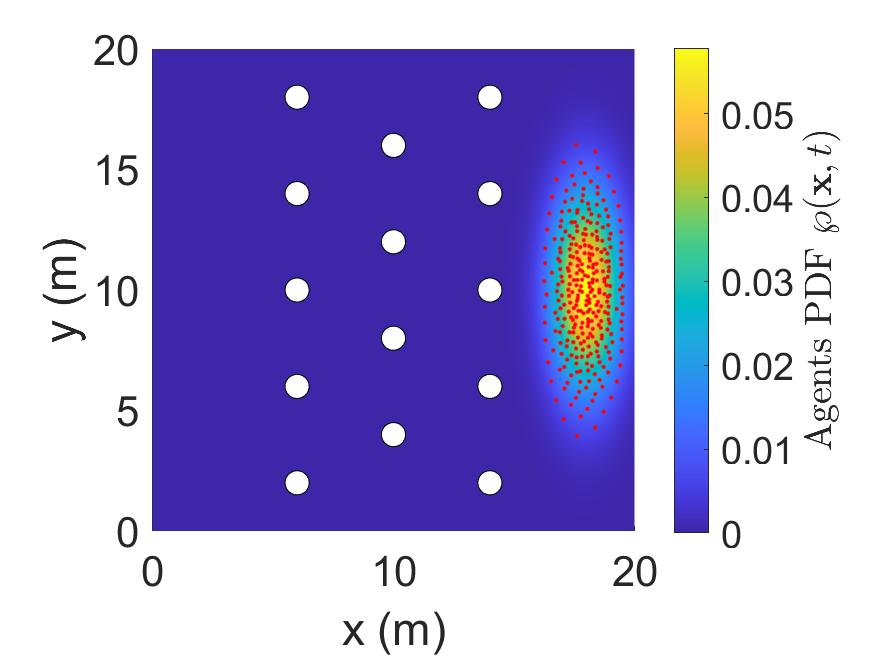} 
	\caption{Snapshots of the trajectory of agents and corresponding PDFs generated
		by the SNMM-APF approach in the Forest-I simulation.  }
	\label{fig:Snapshots_SNMM_APF}
\end{figure}

The SNMM-DI and the SNMM-APF path-planning algorithms are first executed to generate the trajectories of agents' PDFs, where $N_C$ is set to 2, where the SNMMs are also initialized by the GMM's   parameter learning algorithm \cite{peel2000GMM} like Section \ref{sec: Exp-A}. Then, the microscopic controls for individual agents are obtained to manipulate agents to traverse the artificial forest. The snapshots of trajectories of agents and corresponding PDFs generated by SNMM-DI and SNMM-APF path-planing algorithms are presented in Fig. \ref{fig:Snapshots_interpolation} and \ref{fig:Snapshots_SNMM_APF}, respectively. These snapshots show that both path-planning algorithms can provide acceptable performances, although different trajectories of agents' PDFs are generated.   It can also be observed that the time-varying agents' distributions are ``split" by the obstacles due to the skewing function $Q_Y$, which is the expected advantage of the SNMM over the GMM.

To demonstrate the explicit advantage of SNMM over GMM in path-planning problems, a GMM-based approach is also applied to the same problem for comparison.  Given that the GMM can be viewed as a special case of the SNMM where $Q_Y(\mathbf{X}) \equiv 1$ according to (\ref{eq: def_BRF_SNMM}), the APF-based path-planning approach can also be applied to obtain the GMM trajectory by setting $Q_Y(\mathbf{x}) = 1$ for all $\mathbf{x} \in \mathcal{W}$. However, since the obstacle information is not embedded in the GMM distributions, $\varphi(\mathbf{x}, t)$, a repulsive potential against the skewing function is added to the potential defined in (\ref{eq: potential}), such that
\begin{equation}
	U_{Rep}(t) \triangleq \int_{\mathcal{W}} \varphi(\mathbf{x}, t) [1 - Q(\mathbf{x})] d \mathbf{x}
\end{equation}
It is noteworthy that the repulsive potential becomes zero when $Q_Y(\mathbf{x}) = 1$ for all $\mathbf{x} \in \mathcal{W}$, signifying a scenario with no obstacles. The range of the repulsive potential is $U_{Rep} \in [0, 1]$. The potential for the GMM-based approach can be expressed by
\begin{align}
	U_{GMM}(t) \triangleq &  \gamma_{SN} U_{SN}(t \vert Q(\mathbf{x}) \equiv 1) + \gamma_{CS} U_{CS}(t \vert Q(\mathbf{x}) \equiv 1) \nonumber\\
	+& \gamma_{Rep}  \boldsymbol{1}(U_{Rep}(t), \eta) \ln [ U_{Rep}(t) ]
	\label{eq:GMM_potential}
\end{align}
where $\gamma_{Rep} \in \mathbb{R}^{+}$ is a user-defined weight, and $\eta \in [0, 1]$ is a tolerance threshold, which indicates the tolerant maximum cumulated probability of agents over the occupied areas. Finally, similarly to the SNMM-APF approach, the trajectory of agents' PDFs is obtained by minimizing the potential, $U_{GMM}(t)$, w.r.t. the parameters similarly to (\ref{eq: update_mu}) - (\ref{eq: update_Sigma}). For distinction, this GMM-based approach is referred to as the GMM-APF approach.        

\begin{figure}[ht]
	\centering
	\includegraphics[width=\figsize\textwidth]{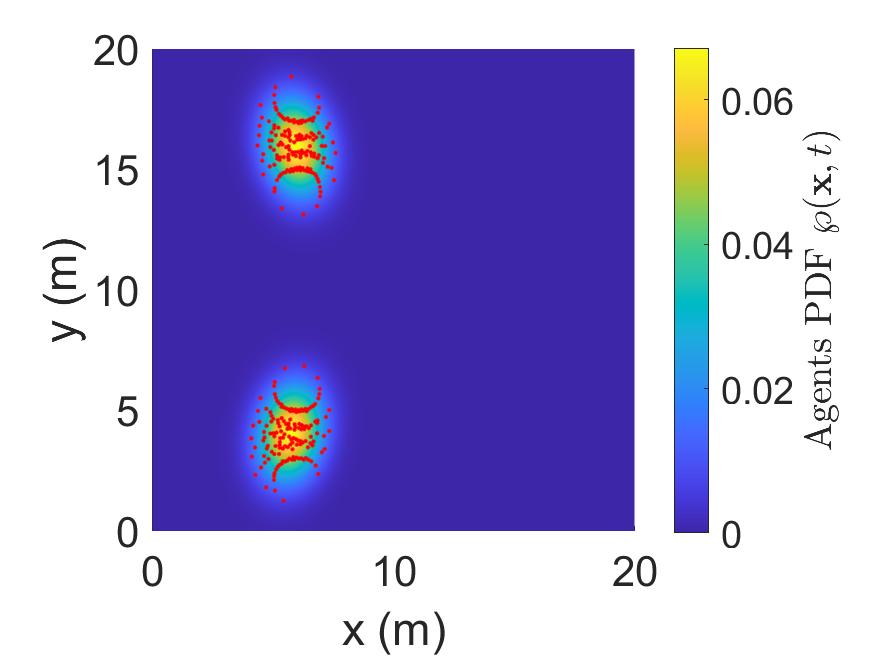} 
	\includegraphics[width=\figsize\textwidth]{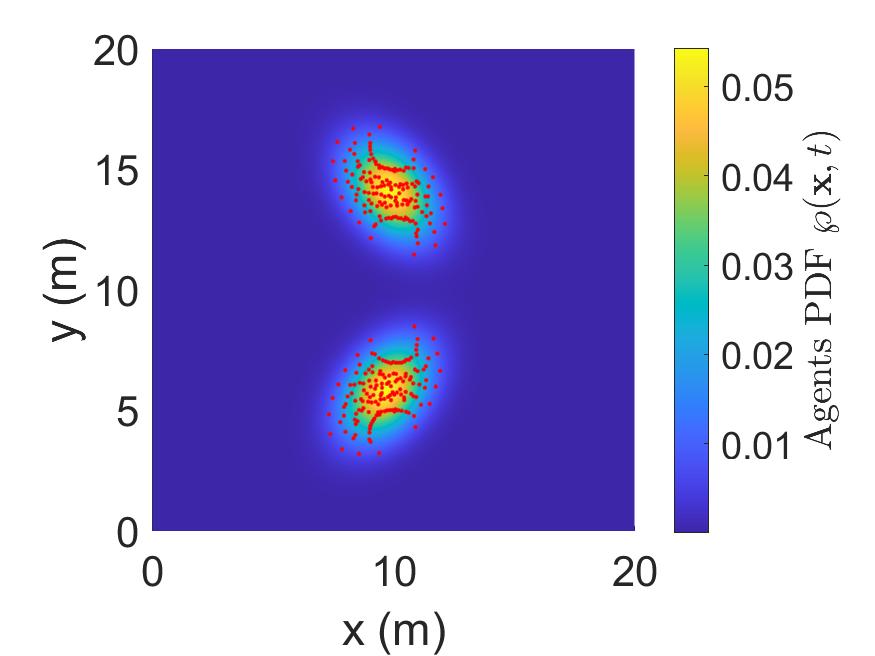} 
	\includegraphics[width=\figsize\textwidth]{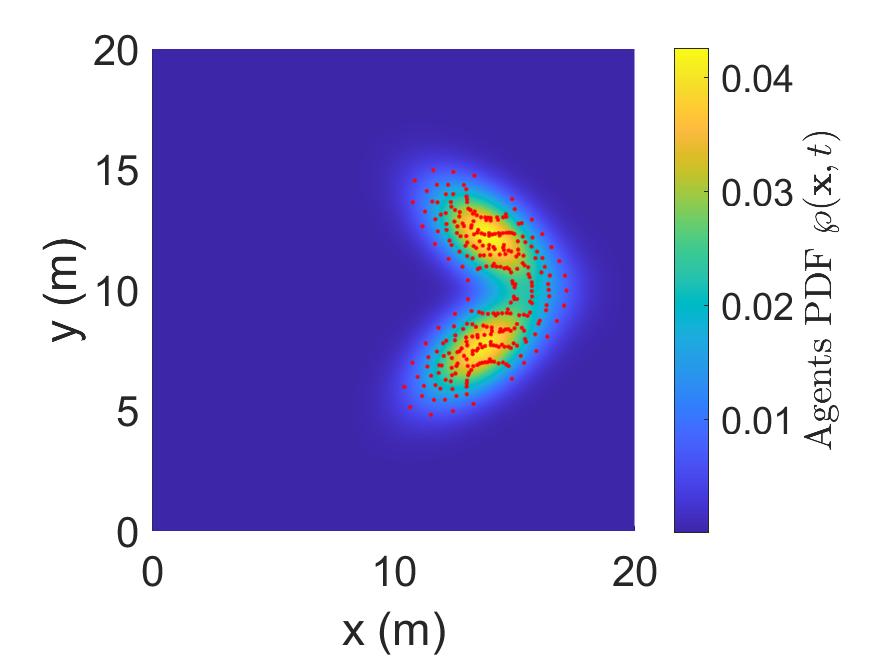} 
	\includegraphics[width=\figsize\textwidth]{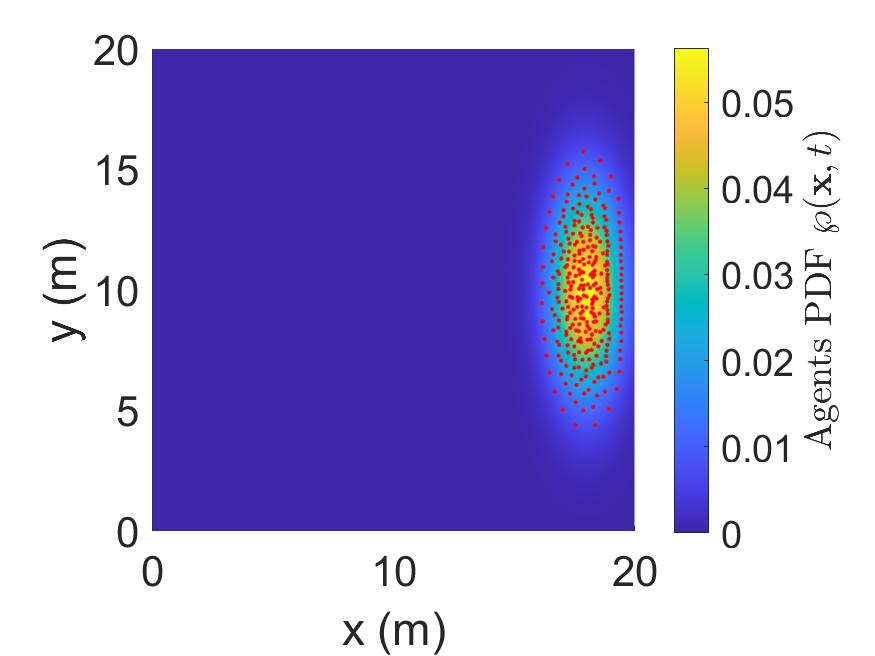} 
	\caption{Snapshots of the trajectory of agents and corresponding PDFs generated
		by the GMM-APF approach in the Forest-I simulation.  }
	\label{fig:Snapshots_GMM_APF}
\end{figure}

For a fair comparison, the same user-defined parameters as the SNMM-APF approaches are applied to the GMM-APF approach. The parameters $\gamma_{Rep}$ and $\eta$ are set to $1$ and $5\%$, respectively, and the max optimization step is set to $3000$. In addition, the same microscopic control is also applied to generate the control inputs. The snapshots of trajectories of agents and corresponding PDFs generated by the GMM-APF algorithm are presented in Fig. \ref{fig:Snapshots_GMM_APF}.

The numerical comparison results are tabulated in Table \ref{tab:numerical_performances}, denoted by ``Forest-I". All of the simulations are conducted by MATLAB on the same computer. The column ``Time" indicates the computational time for obtaining the optimal agents' PDFs. The column ``Length" indicates the statistical results of the length of $N=300$ agents' trajectories in the form of “mean $\pm$ standard deviation”. It can be seen that the proposed two SNMM-based approaches can outperform the GMM-based approach in terms of computational time and trajectory length. In addition, because the SNMM-DI approach does not need to solve the optimization problem, it can obtain the agents' PDFs in a short period. However, it is also noteworthy that the absence of optimization in the SNMM-DI approach will result in a non-continuous PDF trajectory when the VLSR system travels a huge obstacle.

\begin{table}
	\caption{Performance Comparison}
	\label{tab:numerical_performances}
	\centering
	\begin{tabular}{c|c|c|c|c|c}
		\hline 
		            				&				& 	 			&					&								&  				\\	
		Simulation					&	Approach  	& 	Step		& Time  			& Length  						& Success?  	\\
									&				& 	 			& (min)				&	(m)							&				\\	
		\hline
	\multirow{3}{4em}{\textbf{Forest-I}}		&	SNMM-DI 	& \textbf{859} 	&  	\textbf{1.55}	&   \textbf{19.75 $\pm$ 0.92}  	&		Yes		\\ 
			 						&	SNMM-APF 	& 1211 			& 	54.69			&  	27.17 $\pm$ 1.74 			&		Yes		\\ 
									&   GMM-APF 	& 2667 			&  	64.25			&  	33.46 $\pm$ 2.58  			&		Yes		\\ 
		\hline 
	\multirow{3}{4em}{\textbf{Forest-II}}	&	SNMM-DI		& \textbf{859} 			&  	\textbf{1.55}			&  	\textbf{22.68 $\pm$ 1.25}	&		Yes 	\\ 
									&   SNMM-APF 	& 1611 			&  	57.18			&  	30.15 $\pm$ 1.56 			&		Yes		\\ 
									&   GMM-APF 	& 3000 			& 	22.35			& 	18.07 $\pm$ 3.34			& \textbf{No} 	\\
		\hline 
	\end{tabular}
\end{table}

\begin{figure}[ht]
	\centering
	\includegraphics[width=\figsize\textwidth]{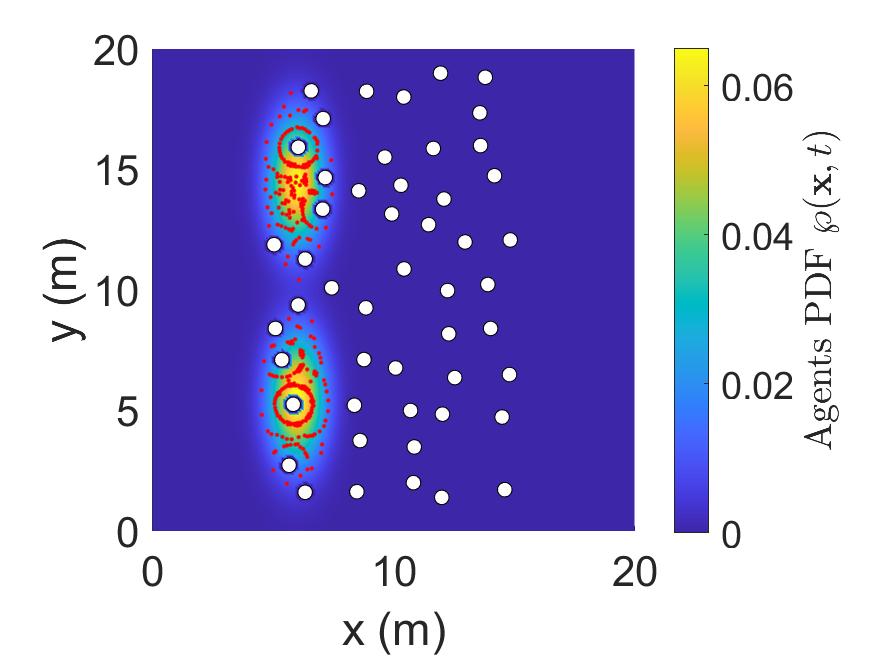} 
	\includegraphics[width=\figsize\textwidth]{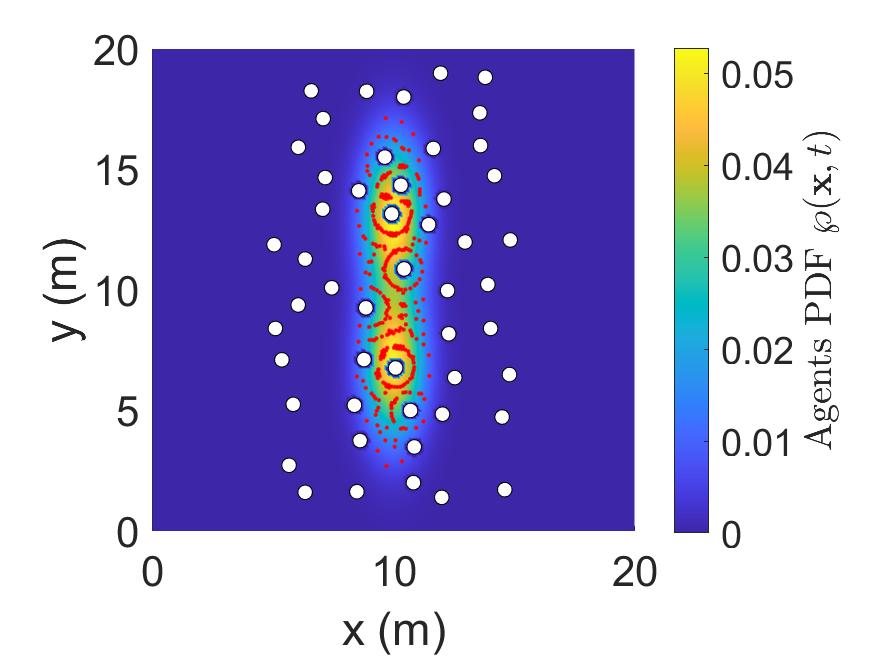} 
	\includegraphics[width=\figsize\textwidth]{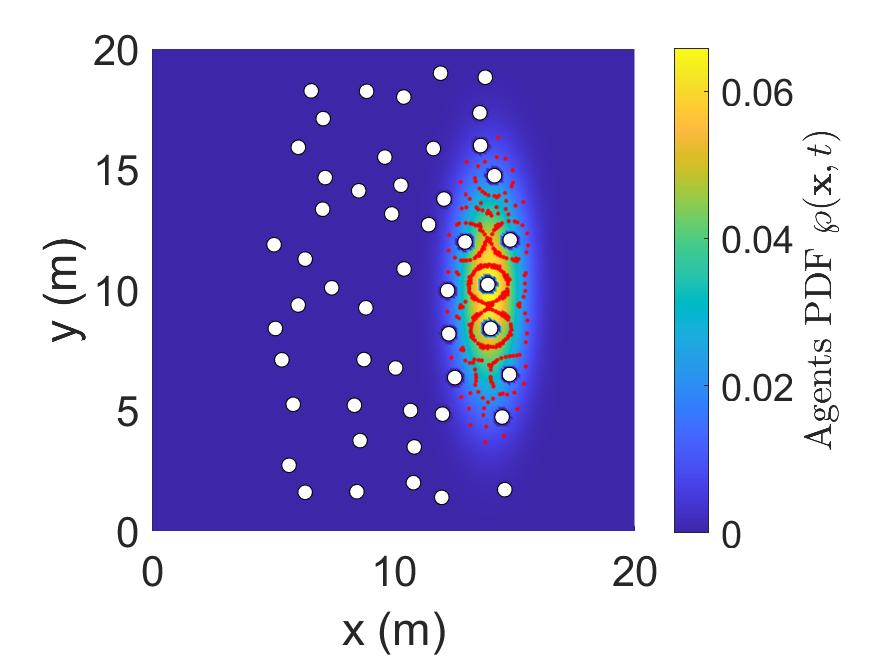} 
	\includegraphics[width=\figsize\textwidth]{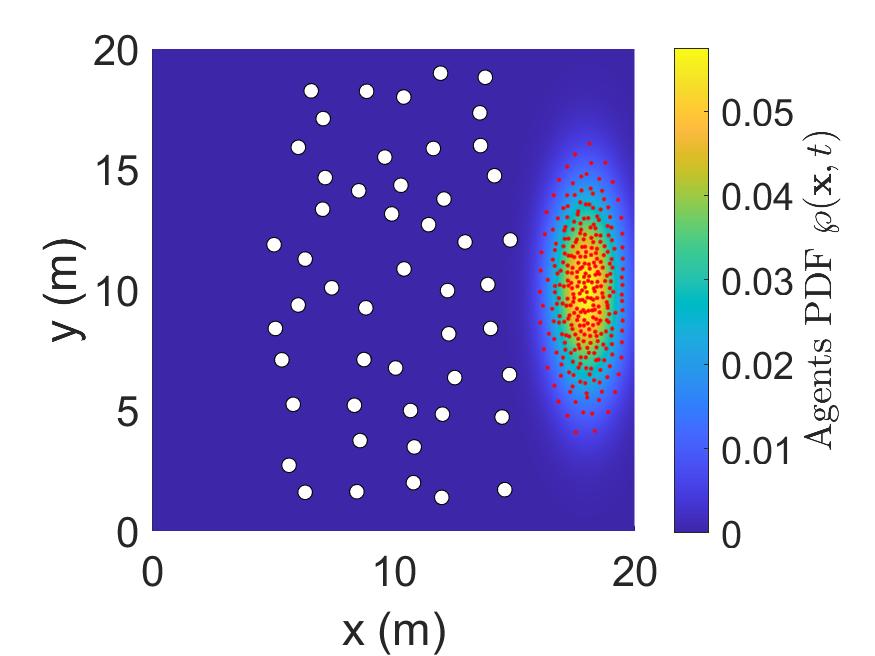} 
	\caption{Snapshots of the trajectory of agents and corresponding PDFs generated by the SNMM-DI approach in the Forest-II simulation. }
	\label{fig:Snapshots_interpolation_Complex}
\end{figure}

\begin{figure}[ht]
	\centering
	\includegraphics[width=\figsize\textwidth]{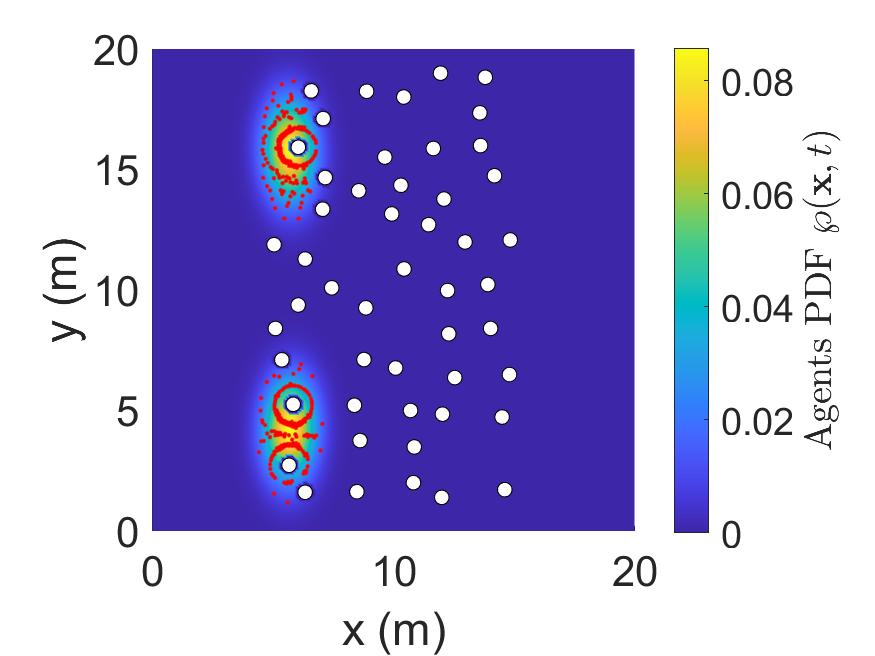} 
	\includegraphics[width=\figsize\textwidth]{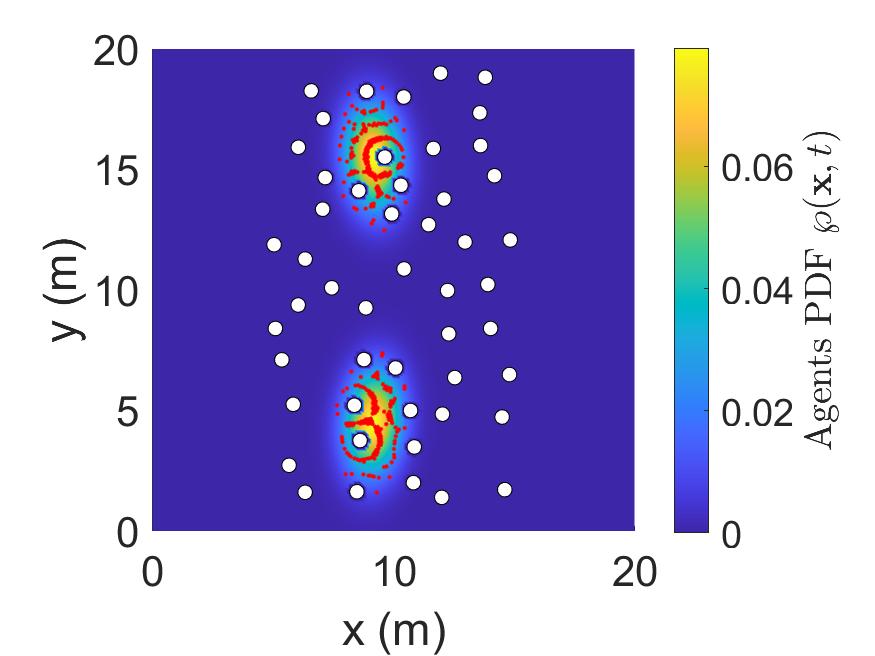} 
	\includegraphics[width=\figsize\textwidth]{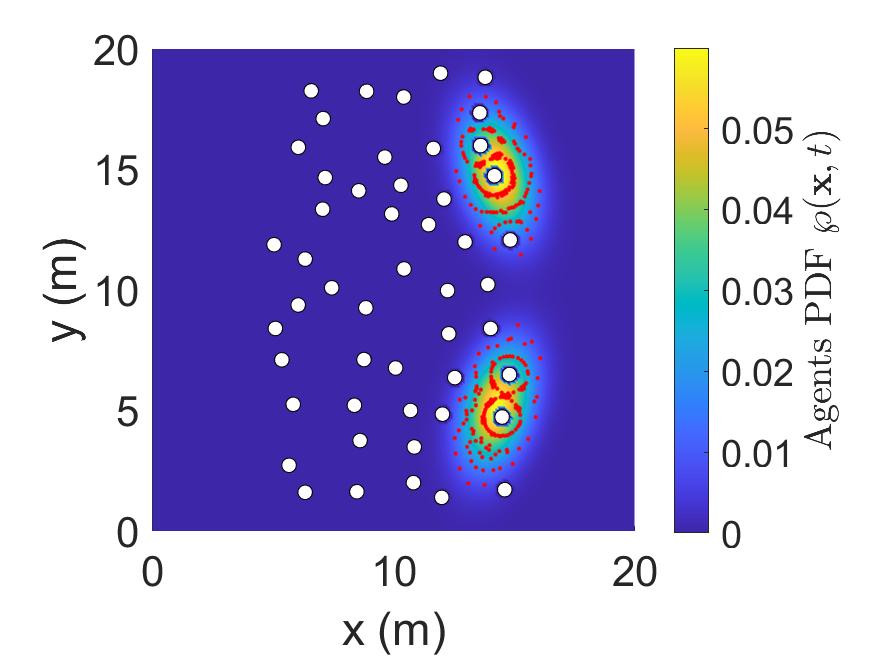} 
	\includegraphics[width=\figsize\textwidth]{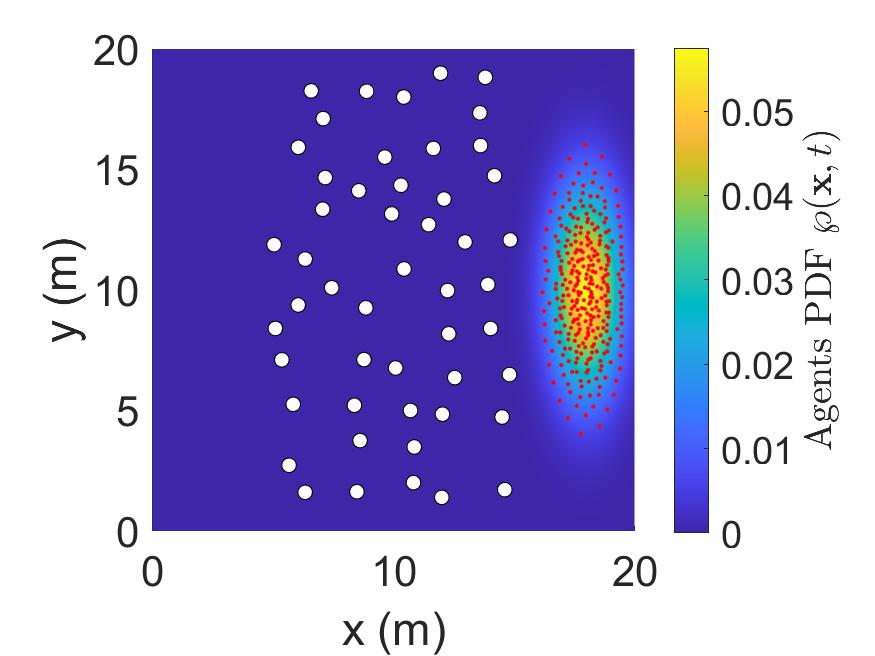} 
	\caption{Snapshots of the trajectory of agents and corresponding PDFs generated
		by the SNMM-APF approach in the Forest-II simulation.  }
	\label{fig:Snapshots_SNMM_APF_Complex}
\end{figure}

\begin{figure}[ht]
	\centering
	\includegraphics[width=\figsize\textwidth]{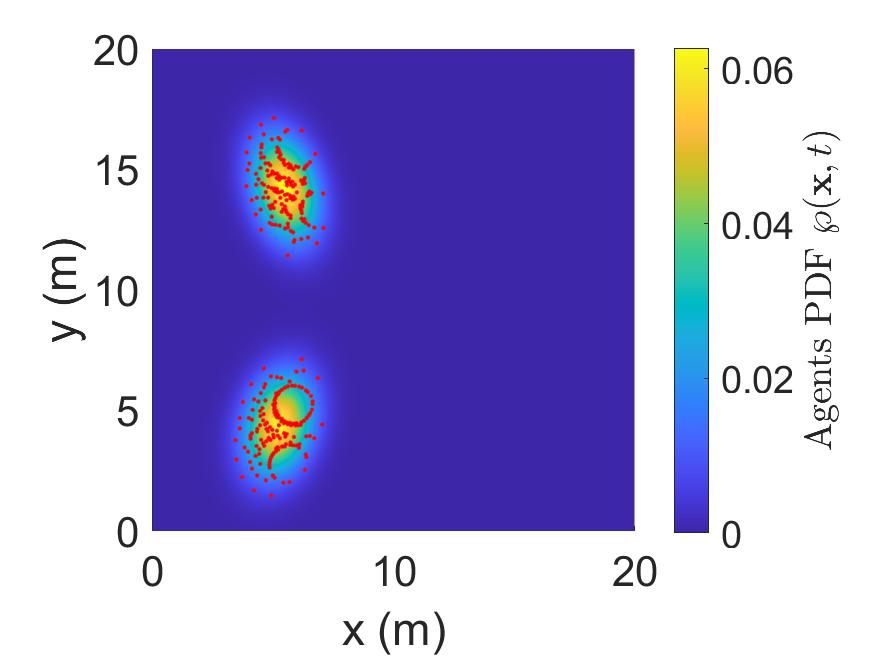} 
	\includegraphics[width=\figsize\textwidth]{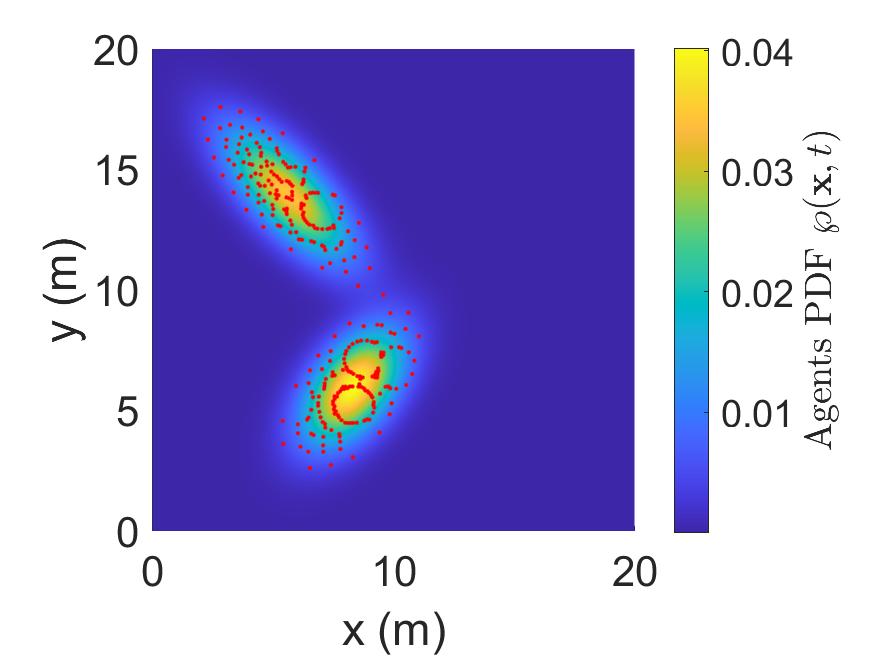} 
	\includegraphics[width=\figsize\textwidth]{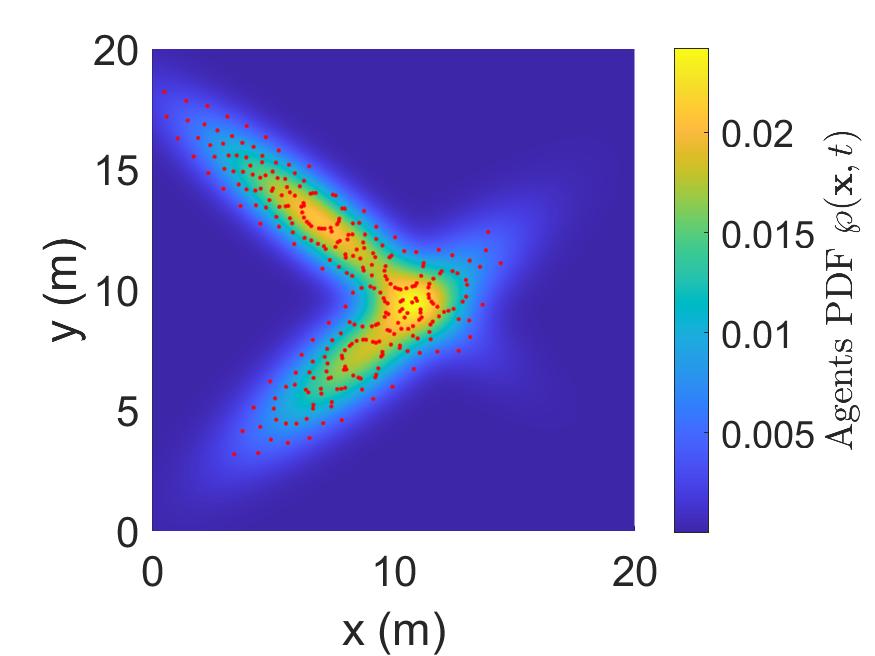} 
	\includegraphics[width=\figsize\textwidth]{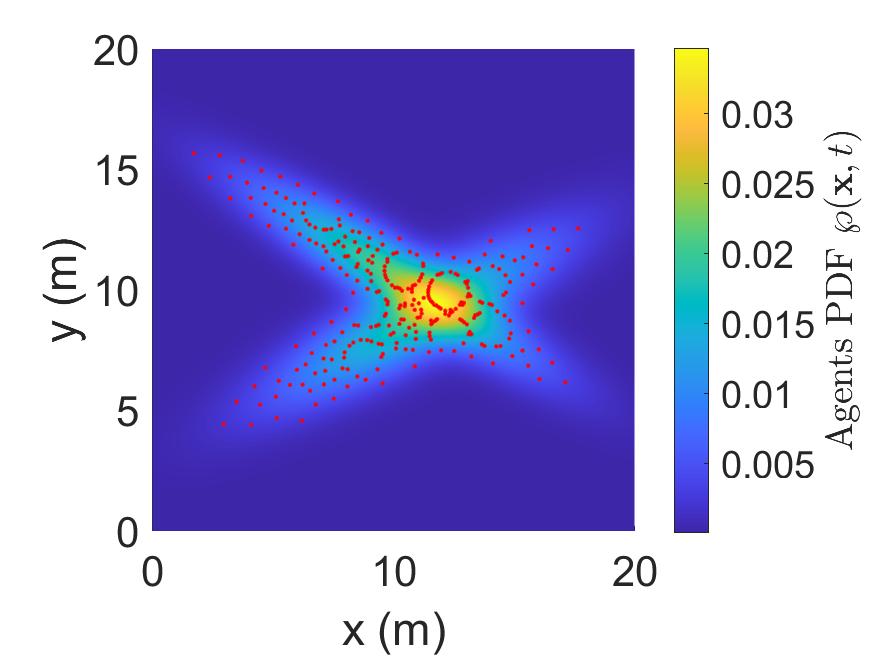} 
	\caption{Snapshots of the trajectory of agents and corresponding PDFs generated
		by the GMM-APF approach in the Forest-II simulation.  }
	\label{fig:Snapshots_GMM_APF_Complex}
\end{figure}

Furthermore, to demonstrate the effectiveness of the proposed SNMM-based approaches for  cluttered environments, a complex artificial forest environment (Forest-II) is also considered, where 50 smaller circles are deployed randomly in the workspace to represent the trees in the forest. Then, assuming the same agents’ initial and the desired distributions, we re-conduct the simulation with the same number of components, $N_C=2$. It is found that our proposed SNMM-based approaches still work well, but the GMM-based approach fails because the potential optimization defined in (\ref{eq:GMM_potential}) is stuck in the local minimum. The agents' trajectories generated by the two SNMM-based and the GMM-based approaches are plotted in Fig. \ref{fig:Snapshots_interpolation_Complex},  \ref{fig:Snapshots_SNMM_APF_Complex}, and \ref{fig:Snapshots_GMM_APF_Complex}, respectively. In addition, the corresponding numerical results are also tabulated in Table \ref{tab:numerical_performances}, denoted by ``Forest-II".        


\begin{figure}[ht]
	\centering
	\includegraphics[width=\figsize\textwidth]{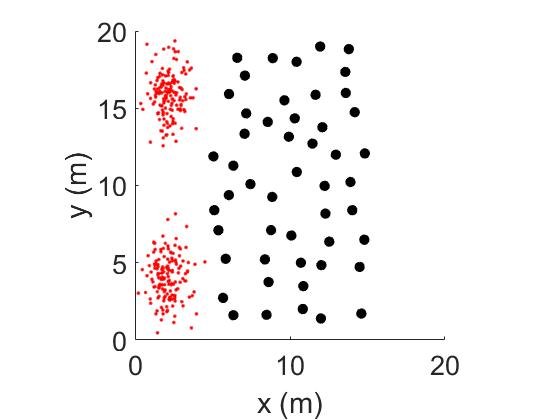}  
	\includegraphics[width=\figsize\textwidth]{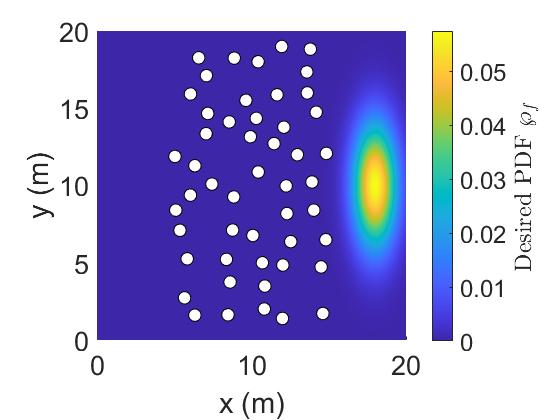} 
	\caption{The path-planning problem for VLSR systems in the complex artificial forest
		environment (Forest-II simulation). Left: The red points indicate the initially deployed agents, and the black circles represent the “trees” in the forest. Right: The desired agents’ PDF is shown, where the white areas are occupied by “trees”.   }
	\label{fig:artifical_forest_II}
\end{figure}

%

\section{CONCLUSIONS AND FUTURE WORK}
This paper proposes a novel Bernoulli-random-field-based multivariate skew-normal distribution and the skew-normal mixture model to represent the macroscopic states of the VLSR systems.  In addition, a parameter learning algorithm is provided to estimate the model parameters for the BRF-SNMM. 
Furthermore, two path-planning algorithms are also developed based on the SNMM to guide the VLSR systems to traverse cluttered environments. The simulations in the artificial forest environments demonstrate the effectiveness of these two path-planning algorithms and the superiority of the SNMM-based approaches over the GMM-based approach, especially in cluttered environments. 







%

\bibliographystyle{IEEEtran}
\bibliography{ref_Ping_skew_normal_distribution}

\end{document}